\documentclass[12pt]{article}

\usepackage{epsfig}
\usepackage[usenames]{color}

\def\be{\begin{equation}}
\def\ee{\end{equation}}
\newcommand{\bea}{\begin{eqnarray}}
\newcommand{\eea}{\end{eqnarray}}
\def\bd{\begin{displaymath}}
\def\ed{\end{displaymath}}

\definecolor{red}{rgb}{1,0,0}

\def\pa{\partial}
\newcommand{\non}{\nonumber \\}
\newcommand{\CR}{\non\cr}
\def\half{\frac{1}{2}}

\begin{document}

\vspace{18pt} \vskip 0.01in \hfill TAUP-290109 \vskip 0.01in \hfill {\tt
hep-th/yymmnnn}

\vspace{30pt}

\begin{center}
{\bf \LARGE   Holographic Technicolor  models and their S-parameter }
\end{center}

\vspace{30pt}

\begin{center}
Oded Mintakevich and  Jacob Sonnenschein

\vspace{20pt}

\textit{School of Physics and Astronomy\\ The Raymond and Beverly
Sackler Faculty of Exact Sciences\\ Tel Aviv University, Ramat Aviv
69978,
Israel\\[10pt]
 }

\end{center}


\begin{center}
\textbf{Abstract }
\end{center}
We study the Peskin-Takeuchi S-parameter of  holographic technicolor models.
We present the recipe for computing the parameter in a generalized holographic setup.
We then apply it to several holographic models that include: (a) the Sakai-Sugimoto model   and (b)  its  non-compactified cousin, (c)  a non-critical analog of (a) based on near extremal $AdS_6$ background, (d) the KMMW model which is similar to model (a) but with  $D6$ and anti-$D6$  flavor branes replacing the $D8$ and anti-$D8$ branes, (e) a model based on D5 branes compactified on two $S^1$s with $D7$ and anti-$D7$ probe branes and (f) the conifold model with the same probe branes as in (e).
The models are gravity duals of gauge theories with $SU(N_{TC})$ gauge theory and with a breakdown of a flavor symmetry $U(N_{TF})\times U(N_{TF})\rightarrow U_V(N_{TF})$. The models (a), (c),(d) and (e) are duals of a confining gauge theories whereas (b) and (f) associate with non confining models.

The S-parameter was found to be S=$sN_{TC}$ where $s$ is given by $0.017\lambda_{TC}$, $0.016\lambda_{TC}$, $0.095$, $0.50$ and $0.043$
for the (a),(b),(c),(d),  (f) models respectively  and for model (e)  $s$ is  divergent. These results are valid in the
large $N_{TC}$ and large $\lambda_{TC}$ limit.
We further derive  the dependence of the S-parameter on the ``string endpoint" mass of the techniquarks for the various models. We compute the masses of the low lying vector  technimesons.


\vspace{4pt} {\small \noindent

 }  \vfill
\vskip 5.mm
 \hrule width 5.cm
\vskip 2.mm {\small \noindent odedm@post.tau.ac.il\\
cobi@post.tau.ac.il}

\thispagestyle{empty}

\eject

\setcounter{page}{1}

\section{Introduction}\label{section1}
One of the most urgent questions in particle physics
is the nature of the mechanism of electro-weak symmetry
breaking (EWSB) and in particular the exact structure of the Higgs sector.
One appealing class of models that may provide an answer to this question
are Technicolor models. In these models a new sector of strongly interacting fermions
known as techniquarks are added to the S.M instead of the scalar Higgs.
This sector will now be responsible for the spontaneous chiral symmetry breaking
($\chi SB$) via a condensate in the TeV scale. The condensate, in a similar
manner to ordinary QCD,  is of a tecniquark anti-techiquark  operator.
The techiquarks transform under certain
representation of the gauge group  $SU(N_{TC})$.
In these models the Higgs boson is a composite state of a
techniquark and an anti-etchniquark  so  that the hierarchy problem is avoided.
One of the most restricting demands of an EWSB model is
that it should produce a small Peskin-Takeuchi S-parameter\cite{Peskin}.
This requirement comes from high precision electro-weak measurement.
The S-parameter defined as\footnote{ For the derivation of the S-parameter its
relation to electroweak measurement and the definition of the variables in this
expression see section (\ref{Peskin}).}
\begin{eqnarray}\label{definition}
S=16\pi[(\Pi'_{33}(0)-\Pi'_{3Q}(0)]
\end{eqnarray}
is restricted to be
in the range of $S=-0.1\pm 0.1$ \cite{Peskin},\cite{Shrock:2007ai} .
A special feature  of the S-parameter that causes it to stand
out among the  high precision measurements, is
that by its very definition, the
S-parameter is isospin-independent. Hence the S-parameter is insensitive to the
exact details of the model by which it is extended to explain the
quarks masses (extended technicolor) and other means
and structures which could be added to the model to explain the
breaking of the isospin symmetry.

The main problem in dealing with technicolor models and in particular in determining their corresponding
S-parameter is the fact that like QCD they are based on a strong dynamics which is non-perturbative
in the region of  interest.
The AdS/CFT correspondence which is by now a very well known
auge/gravity duality provides a useful tool to translate a
strongly coupled gauge system into weakly coupled gravity duals.
This opens the opportunity to use this duality for  describing
technicolor models in terms of weakly coupled gravity.
Indeed the authors of \cite{Carone:2007md} proposed a holographic
technicolor model which is based on the Sakai-Sugimoto model \cite{Sakai:2004cn}. The model is based on Witten's model\cite{Witten:1998zw} of the near extremal limit of $N_{TC}$ $D4$ branes compactified on a circle. Into this background a pair $N_{TF}=2$ of $D8$ and anti-$D8$ branes is incorporated as techiflavor probe branes. In the region that corresponds in the field theory to the UV, the model admits an $U(N_{TF}=2)_L\times U(N_{TF}=2)_R$ chiral symmetry which is spontaneously broken in the IR into a $U(N_{TF}=2)_V$ symmetry. In \cite{Carone:2007md}, using the AdS/CFT dictionary,
the expressions for the axial and vector currents in the boundary field theory where determined. From these one can easily calculate the expressions for the vacuum polarization and their derivatives which then determine the S- parameter. In fact as was discussed in \cite{Carone:2007md} there is a direct way as well as a sum-rule method to compute this parameter.

  The goals of the this paper were threefold, first to generalize the construction  \cite{Carone:2007md}
  to a wide class of
  models that reduces to a five-dimensional effective action of its techniflavor gauge fields.
  Second, to apply this construction to certain concrete holographic models, in particular with or without confinement and spontaneous $\chi SB$
   and  to determine their S-parameter as well as the low lying vector technimesons.
  Third to determine the general properties of holographic models and
  to compare between the holographic estimate of the S-parameter to
  phenomenological estimation of the S-parameter based on scaled up version of QCD.

We start with a five dimensional effective action derived from the DBI  and CS actions of the gauge fields on the probe branes.
Assuming a background that depends only on the radial coordinate, the most general action of this nature was written down.
In analogy to the derivation of \cite{Carone:2007md} the expressions for the S-parameter in both the direct method and the sum-rule method were derived.
This recipe was then applied to the following models:
\begin{itemize}
\item
 (a) The Sakai Sugimoto model \cite{Sakai:2004cn} which is Witten's model
 \cite{Witten:1998zw} generalized to include D8 prob branes and admits confinement and $\chi SB$.
\item
 (b) The uncompactified analog of the latter model. This model, which was analyzed in \cite{Antonyan:2006vw}, is dual of the NJL gauge system which  does not admit confinement but still exhibits spontaneous $\chi SB$.
\item
 (c) A non critical analog of the Sakai Sugimoto model derived in \cite{Kuperstein:2004yk}, \cite{Kuperstein:2004yf}, \cite{Casero:2005se}
 and exhibits  both confining and  spontaneous $\chi SB$. This model consists of $N_{TF}$ pairs of $D4-\!\bar{D}4$ flavor brane probing
 a non critical $AdS_6$ background.
\item
(d) The KMMW model \cite{Kruczenski:2003uq} which is also based on
 Witten's model but with $N_{TF}=2$
 $D6-\!\bar{D}6$  techniflavor probe branes.
 The model admits confinement and a spontaneous breaking of global flavor symmetry which is not a chiral symmetry
\item
 (e) A holographic model \cite{Burrington:2007qd} based
 on the near extremal limit of D5 branes compactified on two circles with $D7\!-\!\bar{D}7$
 flavor barnes. The model is confining and $\chi SB$ is spontaneous.
\item
 (f) The Klebanov-Witten  conifold model with $D7\!-\!\bar{D}7$
 flavor branes \cite{Kuperstein:2008cq}, \cite{Dymarsky:2009cm}.
 This model is conformal before adding the flavor branes but still
 admits spontaneous flavor $\chi SB$.
\end{itemize}

 We found that model (e) fails to serve as a candidate for Techinicolor/Higgs sector
 since it has a divergent S-parameter.
 In  the rest of the  models considered we found a positive S-parameter which is linear in $N_{TC}$
 that is $S=sN_{TC}$.  In models (c), (d) and (f) $s$ is just
 a numerical factor independent  of  $\lambda_{TC}$ where
 $\Lambda_{TC}=g^2_{TC} N_{TC}$. However, in the Sakai Sugimoto model and its AHJK cousin $s$ was found to be  linear in $\lambda_{TC}$ (see table (\ref{table_Of_S})).
  Recall  that the results are valid in the large $N_{TC}$ and large $\lambda_{TC}$
 limit.\\

Holographic Technicolor was studied in recent years also in the following papers \cite{holtech}.
The paper is organized as follows:
We start in section \ref{Peskin} with a brief review of the physics
behind the Peskin-Takeuchi S,T and U parameters. This section is brought for the benefit of the readers which are not familiar with those parameters. Other readers can move directly to section \ref{ThSp}.
Section  \ref{ThSp} is devoted to the determination of the S-parameter in holographic models.
We derive the  formulae for computing the parameter from a general holographic model reduced to five dimensions.
This is done in both a direct method as well as a sum-rule approach.
In section  \ref{SS}  we show the way  the general result  is
implemented   for  the Sakai-Sugimoto model  as was derived in  \cite{Carone:2007md} and we further derive in a qualitative way
the dependence of the S-parameter on the string endpoint masses.
In section \ref{AHJK} we repeat the steps of section  \ref{SS} but for the uncompactified
version of Sakai-Sugimoto model presented by AHJK in \cite{Antonyan:2006vw}.
In section \ref{section3} we examine a  holographic model derived in \cite{Kuperstein:2004yf} which is a non-critical analog of
the Sakai-Sugimoto model
based on  $N_{TF}$ pairs of $D4\!-\!\bar{D}4$ flavor brane probing
a non critical $AdS_6$ background.
Section \ref{section4} is devoted to analyzing the S-parameter of the KMMW model \cite{Kruczenski:2003uq}.
The model is based on Witten's model with
 $D6\!-\!\bar{D}6$  flavor probe branes. In section \ref{section5} we discuss a holographic model \cite{Burrington:2007qd} based
 on the near extremal limit of D5 branes compactified on two circles with $D7\!-\!\bar{D}7$
flavor branes.
We then discuss in section (\ref{section6}) the S-parameter of a model based on the conifold with $D7\!-\!\bar{D}7$
flavor branes. We conclude in section \ref{section7} where
we present the summary of the paper and our conclusion from this work.

\section{Peskin-Takeuchi parameters\cite{Peskin}}\label{Peskin}
In the standard model, the EWSB and fermion masses are explained by
the existence of the Higgs scalar field which acquire a non zero
VEV. The physics of the Higgs sector depends on four
free parameters, the coupling constants
$g'$ and $g$ of the $U(1)_y$  and $SU(2)_L$ respectively,
$v$ the VEV of the Higgs field and its mass $m_h$.
We can express certain observable quantities via these parameters such as
\begin{eqnarray}\label{}
m_W=g\frac{v}{2}\ \ \ ;\  \ \ m_Z=\frac{v}{2}\sqrt{g'^2+g^2}
\ \ \ ;\  \ \  e=\frac{gg'}{\sqrt{g'^2+g^2}}
\ \ \ ;\  \ \   G_f=\frac{1}{\sqrt{2}v^2}
\end{eqnarray}
We used three parameters to define four observable quantities so
there is a hidden relation among them which is independent of the
value of the parameters in the Lagrangian.
Such a relation can be constructed for example by using the
different definition of the weak mixing angle in terms of
observable quantities such as:
\begin{eqnarray}\label{}
s^2=\sin^2\theta_w=\frac{g'^2}{g'^2+g^2}=1-\frac{m_W^2}{m_Z^2}
\end{eqnarray}
\begin{eqnarray}\label{}
\sin2\theta_0=\bigg(\frac{e}{\sqrt{2}G_fm^2_z}\bigg)^{1/2}
\end{eqnarray}
Another useful definition is constructed from the polarization asymmetry
of $Z$ decays into left and right electrons
\begin{eqnarray}\label{}
A_{LR}^e=\frac{\Gamma(Z\rightarrow \bar{e}_Le_L)-\Gamma(Z\rightarrow \bar{e}_Re_R)}
{\Gamma(Z\rightarrow \bar{e}_Le_L)+\Gamma(Z\rightarrow \bar{e}_Re_R)}
=\frac{(\frac{1}{2}-s_*)^2-s_*^2}
{(\frac{1}{2}-s_*)^2+s_*^2}
\end{eqnarray}
These three different definition of the weak mixing angle coincide
at tree level
\begin{eqnarray}\label{tree_level_zero}
\sin^2\theta_w=\sin^2\theta_0=s_*^2
\end{eqnarray}
but their loop corrections are different.
Subtracting them from each other or taking their ratios produces
what is known as $zeroth\  order\ natural\ relation$ which means
a relation which doesn't depends on the parameters of the Lagrangian.
Hence, these relations are free of any UV divergencies coming from
counter terms (since these only alters the parameters of the Lagrangian),
and so the only quantum corrections they receive are finite and can
be considered as predictions of the quantum structure of the theory.
In light of (\ref{tree_level_zero}) we can easily construct
the following zeroth order natural relations:
\begin{eqnarray}\label{zeroth_order_natural_relations}
c^2-c_0^2=s_0^2-s^2\ \ \ ;\ \ \
s_*^2(q^2)-s_0^2\ \ \ ;\ \ \
s^2-s_*^2(q^2)
\end{eqnarray}
where we used the definitions
\begin{eqnarray}\label{}
c^2=1-s^2=\cos^2\theta_w=\frac{m_W^2}{m_Z^2}\ \ \ ;\ \ \ s^2_0=\sin^2\theta_0
\ \ \ ;\ \ \ c^2_0=\cos^2\theta_0
\end{eqnarray}
Another useful zeroth order natural relation we shell use is
the ratio of charged to neutral-current amplitudes denoted
by $\rho_*(0)$, and is equal to one at tree level.
Now, we would like to estimate the radiative corrections to these relations,
and hopefully to divide them into standard model ones and to
those coming from the technicolor sector which supposedly
give the true descriptions of the Higgs sector.
There are many kinds of loop corrections to these zeroth order
natural relations, in addition to the corrections to the vector
boson propagator, there are vertex corrections, box diagrams,
and diagrams with real photon emission.
In strongly interacting technicolor models the techniquarks do not
couple directly to the leptons and at low energies do not appear in the final states,
the only place where
the new physics going to enters is through corrections to the
vector boson propagator via its vacuum polarization
where they appear in loops of techniquark and anti-techniquark pairs.
Otherwise the techniquarks are not observed at low energy, hence these
corrections are called 'oblique'.
We note that in general, loop contributions are not gauge invariant
one by one, but rather their sum is, but since these are the only
contributions involving the techniquarks their gauge invariance is self evident.
As we noted earlier our goal is to sperate the radiative corrections
coming from the new physics from that of the standard model, if we assume
that $m_f\ll m_z$ ($m_f$ is the mass scale of the fermions at the outer legs),
then we can ignore the vertex corrections and box diagrams since these
are suppressed by additional factor of $\frac{m_f^2}{m_Z^2}$ relatively
to the oblique correction.
So we are left with the problem of separating the new physics contributions
to the vacuum polarization from the SM's.
The experimental data that we are trying to fit comes from physics
at energy scales between $\Lambda_{QCD}$ to the $\textrm{TeV}$ scale.
In this range of energies the QCD is weakly coupled
while the techniquarks are still in the strong coupling regime.
Hence we can use perturbation theory to estimate the quarks contributions
to the vacuum polarization amplitude of the gauge fields but we cannot
do so for the techniquarks.
The one loop SM oblique corrections to
(\ref{zeroth_order_natural_relations}) are given by
\footnote{We note that the one loop vacuum polarization
amplitude is proportional to $\frac{m_q^2}{m_Z^2}$ where $m_q$
is the mass of the fermion in the loop, so one only consider the top quark contribution.}
\footnote{Of course one should also consider the contributions coming from
the physical Higgs boson, but since we are replacing this sector
by the technicolor sector, it is omitted \cite{Peskin_book}.}
\begin{eqnarray}\label{standard_model_contributions}
s_*^2-s_0^2&=&-\frac{3\alpha}{16\pi(c^2-s^2)}\frac{m_t^2}{m_Z^2}+\dots
\\ \nonumber
s^2-s_*^2&=&-\frac{3\alpha}{16\pi s^2}\frac{m_t^2}{m_Z^2}+\dots
\end{eqnarray}

Now, denoting $\Pi_{IJ}$ as the 
correlators of the $I$ and $J$ currents of $SU(2)_L\times U(1)_Y$,
where only the contributions coming
from the new physics are taking into account, then
after some algebra we obtain the following form for the radiative corrections
to (\ref{zeroth_order_natural_relations}) due to the Technicolor sector:
\begin{eqnarray}\label{zero_order_corrections1}
c^2-c_0^2&=&s_0^2-s^2=-\bigg[\frac{e^2c^2}{s^2(c^2-s^2)m_Z^2}
\bigg[\Pi_{33}(m_Z^2)-2s^2\Pi_{3Q}(m_Z^2)\\ \nonumber
&-&\frac{s^2}{c^2}\Pi_{11}(0)-\frac{c^2-s^2}{c^2}\Pi_{11}(m_W^2)\bigg]
+\frac{e^2s^2c^2}{c^2-s^2}[\Pi'_{QQ}(m_Z^2)-\Pi'_{QQ}(0)]\bigg]
\end{eqnarray}
\begin{eqnarray}\label{zero_order_corrections2}
s_*^2(q^2)-s_0^2=\bigg[\frac{e^2}{c^2-s^2}\bigg[
\frac{\Pi_{33}(m_Z^2)-2s^2\Pi_{3Q}(m_Z^2)-\Pi_{11}(0)}{m_Z^2}
-(c^2-s^2)\frac{\Pi_{3Q}(q^2)}{q^2}\bigg] \\ \nonumber
+\frac{e^2s^2}{c^2-s^2}[s^2\Pi'_{QQ}(m_Z^2)-c^2\Pi'(0)_{QQ}+(c^2-s^2)\Pi'_{QQ}(q^2)]\bigg]
\end{eqnarray}

\begin{eqnarray}\label{zero_order_corrections3}
\rho_*(0)-1=\frac{e^2}{s^2c^2m_Z^2}
[\Pi_{11}(0)-\Pi_{33}(0)]
\end{eqnarray}

Thus, we see that it is both possible and natural to isolate
the radiative corrections due to new physics from those
coming from the SM fields.
If the new physics included in the vacuum polarization amplitude
is associated with new heavy particles of mass scale $m_{TC}\gg m_Z$,
then we will see a rapid convergence of a Taylor expansion in $q^2$ of
these amplitude. Thus it is natural to expand the $\Pi_{IJ}$ in powers
of $q^2$, neglecting the order $q^4$ and beyond:
\begin{eqnarray}\label{expansion_of_Pi}
\Pi_{QQ}(q^2)&\approx &q^2\Pi'_{QQ}(0)\\ \nonumber
\Pi_{3Q}(q^2)&\approx &q^2\Pi'_{3Q}(0)\\ \nonumber
\Pi_{33}(q^2)&\approx &\Pi_{33}(o)+ q^2\Pi_{33}(0)\\ \nonumber
\Pi_{11}(q^2)&\approx &\Pi_{11}(o)+ q^2\Pi'_{11}(0)
\end{eqnarray}
There are six independent coefficient in (\ref{expansion_of_Pi}) but three
linear combinations of them must cancel out since there are no UV divergences
in (\ref{zero_order_corrections1}), (\ref{zero_order_corrections2}) and
(\ref{zero_order_corrections3}) despite there are in the $\Pi_{IJ}$.
The remaining three are the following:
\begin{eqnarray}\label{S_T_U}
S&\equiv &16\pi[\Pi'_{33}(0)-\Pi'_{3Q}(0)]
\\ \nonumber
T&\equiv &\frac{16\pi}{s^2c^2m_Z^2}\pi[\Pi_{11}(0)-\Pi_{33}(0)]
\\ \nonumber
U&\equiv &16\pi[\Pi'_{11}(0)-\Pi'_{33}(0)]
\end{eqnarray}
Substituting (\ref{S_T_U}) into (\ref{zero_order_corrections1}),
(\ref{zero_order_corrections2}) and (\ref{zero_order_corrections3})
yields
\begin{eqnarray}\label{zero_order_S_T_U}
\frac{m^2_W}{m^2_Z}-c_0^2&=&\frac{\alpha c^2}{c^2-s^2}
\bigg[-\frac{1}{2}S+c^2T+\frac{c^2-s^2}{4s^2}U\bigg]
\\ \nonumber
s_*^2(q^2)-s_0^2&=&\frac{\alpha}{c^2-s^2}[\frac{1}{4}S-s^2c^2T]
\\ \nonumber
\rho_*(0)-1&=&\alpha T
\end{eqnarray}

To summarize, we concluded that under the above assumptions
the dominant radiative corrections to (\ref{zeroth_order_natural_relations})
comes from the vacuum polarization amplitudes, and these
receive contributions from two sources, the standard model part given by
(\ref{standard_model_contributions}) which is fixed and well known,
and a part coming from a sector of new physics 
and are given by (\ref{zero_order_S_T_U}). \\
According to (\ref{zero_order_S_T_U}), we have a three parameter
description of the radiative corrections due to the technicolor sector,
and since the quantities in the left hand side are all observable
there are experimental bounds on the magnitude of these corrections!
In this paper we focus on pure technicolor models without
an extension that could produce isospin violation, in this case
the $T$ and $U$ parameters are zero and we use the
experimental bound on $S$ alone. These experimental bounds restricts S to be
in the range of $S=-0.1\pm 0.1$.

The Peskin-Takeuchi S-parameter  defined above in (\ref{S_T_U}) can be also expressed as
\begin{eqnarray}\label{definition}
S=16\pi[(\Pi'_{33}(0)-\Pi'_{3Q}(0)]=-4\pi[\Pi'_V(0)-\Pi'_A(0)]|
\end{eqnarray}
where $\Pi_V$ and $\Pi_A$ are define by
\begin{eqnarray}\nonumber
i\int d^4x e^{-iqx}\langle \mathcal{J}_{\mu}^{aV}(x)\mathcal{J}_{\nu}^{bV}(0)\rangle
=-\bigg(g_{\mu\nu}-\frac{q_{\mu}q_{\nu}}{q^2}\bigg)\delta^{ab}\Pi_V(q^2)
\end{eqnarray}
\begin{eqnarray}\nonumber
i\int d^4x e^{-iqx}\langle \mathcal{J}_{\mu}^{aA}(x)\mathcal{J}_{\nu}^{bA}(0)\rangle
=-\bigg(g_{\mu\nu}-\frac{q_{\mu}q_{\nu}}{q^2}\bigg)\delta^{ab}\Pi_A(q^2)
\end{eqnarray}
where $\mathcal{J}_{\mu}^{aV}$ and $\mathcal{J}_{\mu}^{aA}$ are the vector current
and axial-vector current respectively.
Using dispersive representation with delta function resonances
these could be expressed by
\begin{eqnarray}\nonumber
\Pi_V(-q^2)=\sum_n\frac{g^2_{V_n}q^2}{m_{V_n}^2(-q^2+m_{V_n}^2)}
\end{eqnarray}
\begin{eqnarray}\nonumber
\Pi_A(-q^2)=\sum_n\frac{g^2_{A_n}q^2}{m_{A_n}^2(-q^2+m_{A_n}^2)}
\end{eqnarray}
It follows then that S could be written as
\begin{eqnarray}\label{sum_rule}
S=4\pi\sum_n\bigg(\frac{g^2_{V_n}}{m_{V_n}^4}-\frac{g^2_{A_n}}{m_{A_n}^4}\bigg)
\end{eqnarray}
where $m_{V_n/A_n}$ and $g^2_{V_n/A_n}$ are the masses and decay constants
of the vector/axial-vector mesons of the confined phase of the Technicolor sector.\\

\section{The holographic S parameter}\label{ThSp}
The general  holographic technicolor  setup  is similar to that of holographic QCD.\footnote{ Below we discuss also
models without confinement or without chiral symmetry}
It is  based on a gravity background that admits confinement in the sense of an area-law behavior of the Wilson line and a discrete spectrum with a mass gap of states dual to the techniglueballs.
The background is characterized by a flux, typically associated with a RR form, denoted here by $N_{TC}$ which corresponds to the rank of the dual technicolor gauge group $SU(N_{TC})$.
A set of $N_{TF}$ flavor probe $D_p$ branes is incorporated in this background.
The worldvolume of the $D_p$  flavor branes includes the four dimensional space-time, the radial direction  and a  $p-4$ non-trivial cycle.
The physics of the  flavor brane is determined   by an action defined on the worldvolume of the probe branes that include a DBI term and a CS term
\be\label{DBICS}
S_{TF}= S_{DBI}+ S_{CS}= -T_p\int d^{p+1}\sigma e^{-\phi}\sqrt{-det(g_{ind} + \cal{F} )} + T_p \int \sum_k C_k\wedge e^{{\cal F}}
\ee
where $T_p$ is the tension of the $D_p$  probe branes,  $g_{ind}$ is the induced metric on those  probe brane, ${\cal F}= 2\pi l_s^2 F + B_{ind}$ where $F$ is the techniflavor field strength associated with
 $U(N_{TF})$ gauge symmetry  and $B_{ind}$ is an induced $B$ field ( if there is one) and $C_k$ is a $k$ RR form.

Since an important ingredient in the technicolor scenario is the spontaneous breaking of the technichiral symmetry, the flavor probe branes have to admit geometrically  in the region dual to the UV flavor chiral symmetry of the form  $U_L(N_{TF})\times U_R(N_{TF})$  and in the IR a spontaneous breakdown of this symmetry to the diagonal subgroup $U_D(N_{TF})$. This requires an  embedding profile of the form of a U shape. Examples of such a holographic setup are the well known Sakai Sugimoto model \cite{Sakai:2004cn}, its non-critical analog  and the recently proposed  model based on incorporating D7 flavor branes in the Klebanov Strassler model.

Integrating the DBI action over the $p-4$ compact cycle expanding in powers of derivatives and gauge fields and  keeping the lowest order one finds the following five dimensional YM action for $U(N_{TF})$ gauge fields.
\be\label{SDBI}
S_{DBI}= -\frac{\kappa_p}{4}\int d^4 x du \left [ a(u) F_{\mu\nu}F^{\mu\nu} + 2b(u)F_{\rho u}F^{\rho u}\right ]
\ee
where $u$ indicates the radial direction, Greek indices are space-time indices,  the contraction of indices is done with $\eta_{\mu\nu}$, $\kappa_p$, $a(u)$ and $b(u)$ are given by
\bea
\kappa_p&\equiv& -\frac{T_p ( 2\pi \alpha') V_{p-4}}{g_s} \CR
a(u) &\equiv& g_s e^{-\phi} \sqrt{det(g_{ind})}(g_{ind}^{\rho\rho})^2 \qquad b(u) \equiv g_s e^{-\phi} \sqrt{det(g_{ind})}g_{ind}^{\rho\rho}g_{ind}^{uu}\CR
\eea
and where we assumed that the induced metric is diagonal and $V_{p-4}$ is the volume of the  compact cycle the probe brane wrap.
The equations of motion associated with the variations of $A_\rho$ and $A_u$ are give by
\bea\label{EOMu}
a(u)\pa^\mu F_{\mu\rho} +\pa^u( b(u) F_{u\rho} &=& 0 \CR
\pa^\rho F_{\rho u} &=& 0 \CR
\eea
As was discussed above the geometrical realization of chiral symmetry implies that the probe is of a form of a U shape with two branches. Thus the profile is a double valued function of the radial coordinate,  which generically  is in the range $\infty\geq u\geq u_0$. It is useful to define a different coordinate $z,\ \infty\geq z\geq -\infty$ so that the boundary  of one branch of the probe brane say the left one,  is at $z=-\infty$ and the boundary of the right one is at $z=+\infty$.
 Expressed in terms of this coordinate the DBI action (\ref{SDBI}) takes the form
\be\label{SDBIz}
S_{DBI}= -\kappa_p\int d^4 x dz \left [ \hat a(z) F_{\mu\nu}F^{\mu\nu} + \hat b(z)F_{\rho z}F^{\rho z}\right ]
\ee
where
\be
\hat a(z) = u'(z) a(u(z)) \qquad \hat b(z) = \frac{ b(u(z))}{u'(z)}
\ee
with $u'(z) =\frac{du}{dz}(z)$.
It is clear that the corresponding equations of motion take the same form as (\ref{EOMu}) with $z$ replacing $u$ and $\hat a$ and $\hat b$ replacing $a$ and $b$. We continue our discussion here using the $u$ coordinates but obviously we can invert the analysis using a $z$ coordinate.

It is convenient at this point to  choose  the $A_u(x,u)=0$ gauge.
The rest of the gauge fields $A_\mu(x,u)$ are expanded in terms  of normalizable non zero modes and non-normalizable zero modes. In addition we divide the gauge fields into vector fields $V_\mu$ which are symmetric  around $u_0$ (or  under $z\leftrightarrow -z$ ) and axial vector fields $A_\mu$ which are antisymmetric.
 Upon further Fourier transforming the space-time coordinate $x^\mu\rightarrow q^\mu$ the expanded fields take the form

\begin{eqnarray}\label{decomposition}
A_{\mu}(q,u)=\mathcal{V}_{\mu}(q)\psi_{V}^0(u)\!+\!\!\mathcal{A}_{\mu}(q)\psi_{A}^0(u)\!+
\!\sum_{n=1}(V_{\mu}^n(q)\psi_{V_n}(u)\!+\!\!A_{\mu}^n(q)\psi_{A_n}(u))
\end{eqnarray}
The normalizable modes are the bulk gauge  fields while the
non-normalizable are by the gauge/gravity  dictionary sources for
boundary currents.
In fact as was shown in \cite{Sakai:2004cn} the gauge transformation that sets $A_u=0$ requires that
the zero modes include massless modes which are the Goldstone bosons associated with the spontaneous breakdown of the techniflavor chiral symmetry.  These modes play obviously an important role in the technicolor mechanism since they will provide the mass of the electroweak gauge bosons once part of the techniflavor symmetry  is gauged.

In terms of this expansion the equations of  motion (\ref{EOMu}) are
\bea\label{EOMpsi}
\frac{1}{ a(u)}\pa^u(b(u)\pa_u\psi_n(u)) &=& -m_n^2\psi_n(u)\CR
\frac{1}{ a(u)}\pa^u(b(u)\pa_u\psi_0(q^2,u)) &=& -q^2\psi_0(q^2,u)\CR
\eea
where we have used for the normalizable modes $\pa^\rho\pa_\rho V_\mu^n=q^2 V_\mu^n= m_n^2 V_\mu^n$ and where we have used  $\pa^\rho V_\rho^n=0$ that follows from the equation of motion. These equations hold for both the vector modes $\psi_{V_n}$ as well as the axial vector modes $\psi_{A_n}$. Note that
the eigenvalue problem
(\ref{EOMpsi}) becomes first order o.d.e for $m^2_n=0$ and so have
only one solution which is the odd one in accordance with the fact that the pions are pseudoscalars.

Plugging the decomposition (\ref{decomposition}) into (\ref{SDBI}) we find
\begin{eqnarray}\label{D4D8_decompos_action}
S_{F^2}&=&-\frac{\kappa_p}{4}\int d^4qdu Tr
\bigg{(} a(u)\bigg{[}\sum_{n=1}\big{[}|F_{\mu\nu}^{Vn}(q)|^2\psi^2_{Vn}(u)+
|F_{\mu\nu}^{An}(q)|^2\psi^2_{An}(u)\big{]}\nonumber \\
&+&
|F_{\mu\nu}^{V0}(q)|^2\psi^2_{V0}(u)+
|F_{\mu\nu}^{A0}(q)|^2\psi^2_{A0}(u))+
2F_{\mu\nu}^{V0}(q)F^{\mu\nu}_{Vn}(-q)\psi_{V}^0(u)\psi_{Vn}(u)\nonumber \\
&+&
2F_{\mu\nu}^{A0}(q)F^{\mu\nu}_{An}(-q)\psi_{A}^0(u)\psi_{An}(u)
\bigg{]}
-2 b(u) \bigg{[}|V_{\mu}^0(q)|^2(\partial_u\psi_V^0)^2
+|A_{\mu}^0(q)|^2(\partial_u\psi_A^0)^2\nonumber \\
&+&
\sum_{n=1}\big{[}|V_{\mu }^n(q)|^2(\partial_u\psi_{Vn})^2
+|A_{\mu }^n(q)|^2(\partial_u\psi_{An})^2\big{]}\bigg{]}
\bigg{)}
\end{eqnarray}
To further reduce the action to four dimensions we have to normalize the $\psi_{Vn}$ and $\psi_{Vn}$ modes.
This is done as follows
Normalizing the gauge field as
\begin{eqnarray}\label{normalization}
\kappa_p\int  du a(u)  \psi_{Vn}\psi_{Vm}=\delta_{nm}
\end{eqnarray}
and the same for $\psi_{An}$.
Had we chosen to use the $z$ coordinates the normalization condition would have same structure with
$\hat a(z)$ replacing $a(u)$.
The resulting   4d YM action reads
\begin{eqnarray}\label{YMaction}
S_{F^2}&=&-Tr\int d^4q
\sum_{n=1}\bigg{(}\frac{1}{4}|F_{\mu\nu}^{Vn}(q)|^2+
\frac{1}{4}|F_{\mu\nu}^{An}(q)|^2
-\half m_{Vn}^2|V_{\mu }^n(q)|^2-\half m_{An}^2|A_{\mu }^n(q)|^2\nonumber \\
&+&\half a_{Vn}F_{\mu\nu}^{V0}(q)F^{\mu\nu}_{Vn}(q)
+\half a_{An}F_{\mu\nu}^{A0}(q)F^{\mu\nu}_{An}(q)
\bigg{)}+S_{source}
\end{eqnarray}
where
\begin{eqnarray}\label{aVn_D4D8}
a_{Vn}=-\kappa_{p}\frac{b(u)}{m_{Vn}^2}\partial_u\psi_{Vn}|_{u=\infty}
\end{eqnarray}
and the same for $a_{An}$ expressed in terms of $\psi_{An}$.
We define $S_{source}$ to be the terms in (\ref{D4D8_decompos_action}) which
involve only the source
\begin{eqnarray}\label{acsource}
\frac{\kappa_p}{2}\int\!\! d^4qdu b(u)
Tr\bigg{\{}
|V_{\mu}^0(q)|^2(\partial_u\psi_V^0)^2
\!\!+\!\!|A_{\mu}^0(q)|^2(\partial_u\psi_A^0)^2\!\!+\!\!|F_{\mu\nu}^{V0}(q)|^2\psi^2_{V0}(u)\!\!+\!\!
|F_{\mu\nu}^{A0}(q)|^2\psi^2_{A0}(u)
\bigg{\}}
\end{eqnarray}
Performing an integration by parts  in the first term  we find
\bea
& & \frac{\kappa_p}{2}\int d^4q Tr \bigg{\{}
|V_{\mu}^0(q)|^2\int du \pa_u[\psi_V^0 b(u) \pa_u\psi_V^0] - \psi_V^0\pa_u[b(u) \pa_u\psi_V^0]\bigg{\}}\CR
&=& \frac{\kappa_p}{2}\int d^4q Tr
|V_{\mu}^0(q)|^2\bigg{\{}[\psi_V^0 b(u) \pa_u\psi_V^0|_{u=u_0}^{u=\infty}+ q^2\int du a(u) (\psi_V^0)^2\bigg{\}}\CR
\eea
The last term cancels exactly the  $|F_{\mu\nu}^{V0}(q)|^2$ term in (\ref{acsource}) and there is a similar cancelation for the axial gauge fields.
Thus the leftover source term takes the form
\begin{eqnarray}
S_{source}=-\frac{1}{2}Tr\int d^4q\bigg{\{}a_{V_0}|V_{\mu}^0(q)|^2+a_{A_0}|A_{\mu}(q)|^2\bigg{\}}
\end{eqnarray}
where
\begin{eqnarray}\label{a_V0_D4D8}
a_{V_0}=-\kappa_pb(u)\partial_u\psi_{V}^0(u,q^2)|_{u=\infty}
\end{eqnarray}
\begin{eqnarray}\label{a_A0_D4D8}
a_{A_0}=-\kappa_{p}b(u)\partial_u\psi_{A}^0(u,q^2)|_{u=\infty}
\end{eqnarray}
where we have used the equations of motion and we have  taken  that $\psi^0|_{u=\infty}=1$ for both the vector and axial vector zero modes.

The coupling between the source $V^0$ $(A^0$) and the vector
(axial) mesons fields can be read from (\ref{YMaction}) after the kinetic
terms of the vector will be diagonalize, this is done
by the transformation
\begin{eqnarray}\label{diag_transformation}
\tilde{V}_{\mu}^n=V_{\mu}^n+a_{Vn}V_{\mu}^0\ \ \ ;\ \ \
\tilde{A}_{\mu}^n=A_{\mu}^n+a_{An}A_{\mu}^0
\end{eqnarray}
Now the action in terms of the new fields is
\begin{eqnarray}\label{D4D8_4d_diag_action}
S_{\tilde{F}}&=&-Tr\int d^4x
\sum_{n=1}\bigg{(}\frac{1}{4}|\tilde{F}_{\mu\nu}^{Vn}(q)|^2
-\frac{1}{2}m^2_{Vn}(\tilde{V}^n_{\mu}-a_{Vn}V_{\mu}^0)\nonumber \\
&+&\frac{1}{4}|\tilde{F}_{\mu\nu}^{An}(q)|^2
-\frac{1}{2}m^2_{An}(\tilde{A}^n_{\mu}-a_{An}A_{\mu}^0)\bigg{)}
+\tilde{S}_{source}
\end{eqnarray}
where
\begin{eqnarray}
\tilde{S}_{source}&=&-\frac{1}{2}Tr\int d^4q\bigg{\{}a_{V0}|V_{\mu}^0(q)|^2
+a_{A0}|A_{\mu}(q)|^2\\ \nonumber
&+&\sum_n\bigg{(}
 a_{Vn}|F_{\mu\nu}^{0V}(q)|^2+a_{An}
|F_{\mu\nu}^{0A}(q)|^2\bigg{)}\bigg{\}}
\end{eqnarray}

and we find that the decay constants are

\begin{eqnarray}\label{decayconst}
g_{Vn}=m^2_{Vn}a_{Vn}=-\kappa_p b(u)\partial_u\psi_{Vn}|_{u=\infty}
\end{eqnarray}
\begin{eqnarray}
g_{An}=m^2_{An}a_{An}=-\kappa_p b(u) \partial_u\psi_{An}|_{u=\infty}
\end{eqnarray}

Now we have assembled all the ingredients to determine the value of the holographic S parameter.
As discussed in the previous section, this can be done in two different ways.  In the first method we need to compute holographically the two point functions of the vector and axial vector currents.
Using the AdS/CFT dictionary this reads
\begin{eqnarray}\label{AdS/CFT}
-\Pi_V(q^2)\equiv\langle \mathcal{J}_V^{\mu }(q^2)\mathcal{J}_V^{\nu }(0)\rangle_{F.T}=
\frac{\delta}{\delta V^{\nu }_0}\frac{\delta}{\delta V^{\mu }_0}S_{DBI}|_{V^0=0}= a_V^0(q^2)
\end{eqnarray}
where $V^0_{\mu}$ is the boundary value of the vector gauge field at $u=\infty $. The same applies also for the axial vector correlator.
Substituting (\ref{AdS/CFT}) into (\ref{definition})
 the holographic S-parameter reads
\begin{eqnarray}\label{Holographic_S_in_General}
S&=&-4\pi(\Pi'_V(q^2)-\Pi'_A(q^2))|_{q^2=0}=-4\pi\frac{\partial}{\partial q^2}(a_V^0(q^2)-a_A^0(q^2))|_{q^2=0}\CR
&=& -4\pi \kappa_p \left [ b(u)\frac{\partial}{\partial q^2}(\pa_u(\psi_V^0(u,q^2)-\pa_u\psi_A^0(u,q^2))\right]_{q^2=0;u=\infty}
\end{eqnarray}

The second method is based on inserting the decay constants (\ref{decayconst}) into the expression
for the S parameter as sum over resonance given in (\ref{sum_rule}), yielding
\be\label{sum_rule}
S= 4\pi\sum_n\left [ (a_{V_n})^2-(a_{A_n})^2\right ] = 4\pi (\kappa_p) b^2(u)\sum_n\left [(\pa_u\psi_{V_n})^2-(\pa_u\psi_{A_n})^2\right ]_{u=\infty}
\ee
Here we have used the gauge/gravity duality rules and derived the holographic form of the two expressions (ref{definition}) and  (\ref{sum_rule}) that were shown  in the boundary field theory to be equivalent. In fact one can show directly in the gravity
setup that the two expressions are equivalent. This was done in \cite{Carone:2007md} for the sakai Sugimoto model but can be done in a similar way for the general setup discussed in this section. The issue of when a partial sum of a small number of low lying states is a good approximation to the full sum is discussed in \cite{Hirn:2007bb}.

The determination of the $S$-parameter follows from the solutions of the equations of motion
(\ref{EOMpsi}). The latter, as will be seen in the following sections, depend on the profile of the probe brane and in particular on the point with minimal value of the radial direction $u_0$. This parameter relates to the ``string endpoint mass" of the meson ( technimesons in our case) which are defined as follows \cite{Kinar:1998vq}
\be\label{sepm}
m_{sep}=\frac{1}{2\pi\alpha'}\int_{u_\Lambda}^{u_0}\sqrt{-g_{tt}g_{uu}}du
\ee
This mass is clearly not the current algebra or QCD mass, and in fact it is also not the constituent mass of the meson. This mass can be thought of as $m_{sep}=\frac12(M_{meson}-T_{st} L_{st})$ where $T_{st}$ is the string tension and $L_{st}$ is the length of the string.
The fact that it is not the QCD mass is easily determined from the fact that the pions associated with a probe brane profile with non trivial $u_0$ are massless. Thus this mass parameter is not related at all to the masses of particles running in the loops that determine the $S$ parameter. Hence we should not expect the dependence of the $S$ parameter to resemble that of the dependence of the QCD masses. Indeed as will be seen in the sections below the dependence on $m_{sep}$ or on $u_0$  will be different in the various models studied and nor related the dependence on the QCD masses.


\section{The Sakai Sugimoto model}\label{SS}
The starting point of the hologrphic Technicolor Sakai Sugimoto model
is  Witten's model \cite{Witten:1998zw}. The model describes the near extremal  limit of $N_{TC}$ $D4$-branes
wrapping a circle in the $x_4$ direction with anti periodic
boundary condition for the fermions. Having in mind the use of the model as a hologrphic technicolor model, we use from the onset $N_{TC}$ and below $N_{TF}$ instead of $N_c$ and $N_f$ of the original model.
In order to incorporate
fundamental quarks in this model it was suggested in \cite{Sakai:2004cn}
to add to this background a stack of $N_{TF} \ D8$ branes  and a stack of  $N_{TF}$ anti  $D8$ brnaes
Assuming $N_{TF}<<N_{TC}$  the backreaction of the flavor probe branes
 can be neglected as was shown to leading order in $\frac{N_{TF}}{N_{TC}}$ in \cite{Burrington:2007qd}.
The background which includes the metric the RR form and the dilaton is given by
\begin{eqnarray}\label{S&S_metric}
ds^2\!\!\!\!&=&\!\!\!\bigg( \frac{u}{R_{D4}}\bigg)^{3/2}\!\bigg[\!\!-\!\!dt^2\!+\!\delta_{ij}dx^idx^j+f(u)dx_4^2\bigg]
\!+\!\bigg( \frac{R_{D4}}{u}\bigg)^{3/2}\!\bigg[\frac{du^2}{f(u)}\!+\!u^2d\Omega_4^2\bigg]\\ \nonumber
F_4\!&=&\!\frac{2\pi N_c}{V_4}\epsilon_4\ \ ,\ \ e^{\phi}=g_s\bigg( \frac{u}{R_{D4}}\bigg)^{3/4}
 , \ R_{D4}^3=\pi g_sN_cl_s^3 \ , \  \ f(u)=1-\bigg( \frac{u_{\Lambda}}{u}\bigg)^3
\end{eqnarray}
where $V_4$ denotes the volume of the unit sphere $\Omega_4$ and $\epsilon_4$ its corresponding
volume form. $l_s$ is the string length and $g_s$ is the corresponding  string coupling.
The techniflavor branes are placed in such a way that  the compactified
 $x_4$  direction is transverse to them  asymptotically. The manifold spanned by the coordinate $u,x_4$ has the topology of a cigar where its tip is at the minimum value of $u$ which is $u=u_{\Lambda}$.
The periodicity of this cycle is uniquely determine to be
\begin{eqnarray}
\delta x_{4}=2\pi R=\frac{4\pi}{3}\bigg( \frac{R_{D4}^3}{u_{\Lambda}}\bigg)^{1/2}=2\pi R
\end{eqnarray}
in order to avoid a conical singularity at the tip of the cigar.
We also see that the typical scale of the glueball masses computed from
excitation around (\ref{S&S_metric}), is
\begin{eqnarray}
M_{gb}=\frac{1}{R}
\end{eqnarray}
The confining string tension in the model is given by\cite{Kinar:1998vq}
\begin{eqnarray}\label{stringtension}
T_{st}=\frac{1}{2\pi\ell_s^2}\sqrt{g_{xx}g_{tt}}|_{u=u_{\Lambda}}
=\frac{1}{2\pi\ell_s^2}\bigg( \frac{u_{\Lambda}}{R_{D4}}\bigg)^{3/2}
\end{eqnarray}

Corresponding to $u_\Lambda$ one defines   the following mass scale
\begin{eqnarray}
M_{\Lambda}=\frac{1}{R}=\frac{3}{2}\frac{u_{\Lambda}^{1/2}}{R^{3/2}_{D4}}
\end{eqnarray}
Naively one could assume that at energies below $M_{\Lambda}$,
the dual gauge theory is effectively four dimensional; however
since the theory confines and develops a mass gap of order $M_{gb}\sim M_{\Lambda}$
there is no real separation in mass  between the confined four dimensional hadronic  modes,
like the glueballs
and  the Kaluza-Kleine excitations on the $x_4$ circle.
As discussers in \cite{Witten:1998zw} in the opposite limit where
$\lambda_5=g^2_5N_c\ll R$ one can see from loop calculations that
the scale of the mass gap is exponentially small compared to $1/R$
hence the theory does approach the $3+1$ pure Yang-Mills theory
at low energies. It is believed that there is no phase transition
when varying $\lambda_5/ R$ interpolating between the gravity regime to
pure Yang-Mills.
For convenience we will use from here on the freedom to re-scale
the $u$ coordinate and set $u_{\Lambda}=1$.

The flavor probe brane  are space filling  in all the direction
except on the cigar where we need to find their classical curve.
In this case the problem is reduce to an o.d.e for $x_4=x_4(u)$
that follows from the equation of motion associated with  the DBI action of the $D8$ branes.
In fact the general form of the profile can be determined even without the equations of motion. In the geometry of the cigar the flavor branes cannot end and hence they have to fold back  and end asymptotically  at $u\rightarrow\infty$ again transverse to the $x_4$ direction.
The solution  of the equation of motion is found to be
\begin{eqnarray}\label{u_Lamabda_ne_u}
x_{4}(u)=\int^{u}_{u_0}
\frac{du}{f(u)(\frac{u}{R_{D4}})^{3/2}\sqrt{\frac{f(u)u^8}{f(u_0)u_0^8}-1}}
\end{eqnarray}
where $u_0$ is a constant of integration which determines the  lowest value of $u$
to which the $D8$ branes  extend to before folding back  to the UV. Notice that this
U shape with a tip at $u=u_0$ generalizes 
the model of \cite{Sakai:2004cn}. The interpretation of $u_0$, as the string endpoint mass was
discussed in section (\ref {ThSp}).
Since the orientation of the $D8$ was flipped while passing
through $u_0$ it is actually  a $\bar{D}8$ brane and so we have
$D8\!-\!\bar{D}8$ system. As was mentioned above a more natural way to look at this situation is
that one begins with $N_{TF}$ $D8$ located for $u\rightarrow \infty$ at $x_4=x_4^{(L)}$ and $N_{TF}$ $\bar{D}8$ brane at $x_4=x_4^{(R)}$
and finds that  due to the classical equations of motion they  join together at $u=u_0$.
In the  model of  \cite{Sakai:2004cn}  $x_4^{(L)}=0$ and  $x_4^{(R)}=\pi$.
We see that the global
$U_L(N_{TF})\times U_R(N_{TF})$ chiral symmetry of the theory is spontaneously broken
by the ground state down to $U_V(N_{TF})$. So, we got a  gravity model whose
dual gauge theory admits  at low energies confinement and chiral symmetry
is spontaneously broken.
These two qualities are of great importance in QCD phenomenology
and also in building technicolor models.
An HTC model means we identify the gauge group as
the technicolor $SU(N_{TC})$ and the quarks are techniquarks and
the vector field fluctuation of the $D8$ branes are vector technimesons.

We now  compute  S-parameter associated with the technicolor model based on the generalized  Sakai Sugimoto model by applying the two methods described in section (\ref{ThSp}).
The action of the gauge fields is given by
\begin{eqnarray}\label{gaug_DBI_D4D8}
S_{F^2}=-\frac{T_8(2\pi \alpha')^2 V_{S_4}R^{9/2}_{D4}u_{\Lambda}^{1/2}}{4g_su_{\Lambda}^{1/2}}
\int d^4xdu\bigg{\{}\bigg{(}\frac{\gamma}{u}\bigg{)}^{1/2}F_{\mu\nu}^2+\frac{2u^{5/2}}{R_{D4}^3\gamma^{1/2}}F_{\mu u}^2\bigg{\}}
\end{eqnarray}
Using the general discussion of section  (\ref{ThSp}), the model is characterized by
\bea\label{SS_charactization}
\kappa_{8}&=& \frac{T_8(2\pi \alpha')^2 V_{S_4}R^{9/2}_{D4}u_{\Lambda}^{1/2}}{g_s}=\frac{g^2N^2}{36\pi^2} \CR
a(u)&=&(\frac{\gamma}{u})^{\frac12} \qquad b(u)=\frac{u^{5/2}}{R_{D4}^3\gamma^{1/2}} \CR
\eea
where
\be
\gamma= \frac{u^8}{f(u)u^8-f(u_0)u_0^8}
 \ee
Solving numerically equations (\ref{EOMpsi}) for the present case and plugging the results into (\ref {definition}) we reproduced the results of \cite{Carone:2007md}.
 For the
anti-podal configuration $(u_0=u_{\Lambda}=1)$
\begin{eqnarray}
S=19.66\kappa_{D8}= 19.66 \frac{g^2 N^2}{36\pi^3}= 0.017 \lambda_{TC} N_{TC}
\end{eqnarray}
where $\lambda=g^2 N_{TC}$.
The authors of  \cite{Carone:2007md}, used the values $N_{TC}=4$ and $ \lambda_{TC}=4\pi$  to   compare the holographic computation with  the results of \cite{Peskin}. This kind of comparison has to be taken with a grain of salt since the holographic result is valid in the limit of large $\lambda_{TC}$ and large $N_{TC}$.

For the general non anti-podal configurations we find that
S is growing linearly with $u_0$.
Another useful results could be obtained from Weinberg sum rule
\begin{eqnarray}\label{}
\Pi_A(0)=F_{\pi}^2=(246\textrm{GeV})^2
\end{eqnarray}
where we assigned the thechni-pion decay constant the value
of the electroweak scale in order to reproduce the spectrum
of the electroweak gauge bosons.
Using (\ref{a_A0_D4D8}) this gives
\begin{eqnarray}\label{}
F_{\pi}^2=a_{A0}(0)=-\kappa_{D8}R_{D4}^{-3}u^{5/2}\gamma^{-1/2}
\partial_u\psi_{A}^0(u)|_{u=\infty}
\end{eqnarray}
Numerical integration of (\ref{EOMpsi}) gives
\begin{eqnarray}\label{F_pi_D4D8}
F_{\pi}^2=a_{A0}(0)=0.42\kappa_{D8}M_{KK}^2=0.019M_\Lambda^2
\end{eqnarray}
where like in \cite{Carone:2007md} we choused to use the
values $\lambda_{TC}=4\pi $ and $N_{TC}=4$.
This set the Kaluza-Klein mass scale to $M_\Lambda=1.8\textrm{TeV}$ and we can determine now
the mass spectrum of the mesons. We find the first vector technimeson
resonance to be $m_{\rho}\approx 1.5\textrm{TeV} $ and the techni-axial
vector meson $m_{a_1}\approx 2.2\textrm{TeV} $ \cite{Carone:2007md}.
Actually, after the assignment of the Kaluza-Klein scale there are
no more free parameter in the theory except $u_0$ which has no
$4d$ interpretation.\\
As we pointed out in the previous section, there is another way to
to estimate the S-parameter by using
the sum over hadronic resonance given in (\ref{sum_rule}).
Summing up to $n=8$  we find
\begin{eqnarray}
S_8=-1\kappa_{D8}\approx-.001\lambda_{TC}N_{TC}
\end{eqnarray}
Thus the contribution from the eight lowest states is negative
and very far from the result found above.
This is in accordance with the  statement made in section
based on \cite{Hirn:2007bb}, that the higher KK modes do
not decouple from the spectrum on this background
and that $S_n$ for some finite $n$ does not produce a good approximation for $S$.

Next we want to study the dependence of  the S-parameter on
$u_0$. Using (\ref{Holographic_S_in_General}) for different values of
$u_0$ we found  numerically
that $S$ tends to grow linearly  with $u_0$.  We
will  see this behavior in a more qualitative manner for large $u_0$ by using
scaling argument on either one of (\ref{Holographic_S_in_General}) or (\ref{sum_rule}).
We will show how to apply this argument on the scheme given in (\ref{sum_rule})
but we note that it could be applied easily the same to (\ref{Holographic_S_in_General}).
We start by changing variable in (\ref{EOMpsi})
to $y=\frac{u}{u_0}$ and then we take the limit $u_0>>1$, in this limit
we find that

\begin{eqnarray}\label{gamma_D8}
\gamma\to \tilde{\gamma}(y)=\frac{y^8}{y^8-1}
\end{eqnarray}
In this limit the action of the gauge field in (\ref{gaug_DBI_D4D8})
could be written as
\begin{eqnarray}\label{gaug_DBI_D4D8_y}
S_{F^2}=-\kappa_8 \bigg(\frac{u_0}{u_{\Lambda}}\bigg)^{1/2}
\int d^4xdy\bigg{\{}\bigg{(}\frac{\tilde{\gamma}}{y}\bigg{)}^{1/2}F_{\mu\nu}^2+2\frac{u_0}{R_{D4}^3}\frac{y^{5/2}}
{\tilde{\gamma}^{1/2}}F_{\mu y}^2\bigg{\}}
\end{eqnarray}
and (\ref{EOMpsi}) becomes
\begin{eqnarray}\label{eom_for_scaled_psi_n}
y^{1/2}\tilde{\gamma}(y)^{-1/2}\partial_y(y^{5/2}\tilde{\gamma}(y)^{-1/2}
\partial_y\tilde{\psi_{n}}(y))=-\frac{m_n^2R_{D4}^3}{u_0}\tilde{\psi}_n(y)
\end{eqnarray}
Since the left hand side of (\ref{eom_for_scaled_psi_n}) is independent
of $u_0$, then so is the right one, and we find
\begin{eqnarray}\label{qualtative_m_D4D8}
m^2_{Vn/An}\sim \frac{u_0}{R_{D4}^3}
\end{eqnarray}
Now doing the same manipulation on (\ref{decayconst}) we get
\begin{eqnarray}\label{scaled_decay_const_D4D8}
g_{Vn}=-u_0^{3/2}\kappa_{D8}R^{-3}y^{5/2}\tilde{\gamma}^{-1/2}\partial_y\psi_{Vn}(y)|_{y=\infty}
\end{eqnarray}
we might conclude that
\begin{eqnarray}\label{qualitative_decay_const_D4D8}
g_{Vn/An}\sim \frac{u_0^{3/2}}{R_{D4}^3}
\end{eqnarray}
but since the normalization of the modes (\ref{normalization}), is now taken to be:
\be
\kappa_8\bigg(\frac{u_0}{u_{\Lambda}}\bigg)^{1/2}\int dy
\bigg{(}\frac{\tilde{\gamma}}{y}\bigg{)}^{1/2}\tilde{\psi}_n(y)\tilde{\psi}_m(y)=\delta_{mn}
\ee
we need to take into account that
\be\label{rescaled_psi}
\psi(u)=u_0^{-1/4}\tilde{\psi(y)}
\ee
Combining (\ref{scaled_decay_const_D4D8}) and (\ref{rescaled_psi}) we find
\begin{eqnarray}\label{qualitative_decay_const_D4D8_2}
g_{Vn/An}\sim \frac{u_0^{5/4}}{R_{D4}^3}
\end{eqnarray}
Now plugging (\ref{qualtative_m_D4D8}) and (\ref{qualitative_decay_const_D4D8_2})
into (\ref{sum_rule}) we find
\begin{eqnarray}\label{qualitative_Sn_D4D8}
S_n\sim u_0^{1/2}
\end{eqnarray}
\section{ The AHJK model- the uncompactified Sakai Sugimoto model}\label{AHJK}
In the Sakai Sugimoto model the spontaneous breaking of the techniflavor symmetry is attributed to the U-shape configuration of the $D8-\!\bar{D}8$ branes. As was mentioned above this is a result of the cigar structure of the submanifold of the $(u, x_4)$ directions. It turns out that this is a sufficient condition for having a U-shape form but it is not a necessary condition. That is to say that there is a U shape solution even if $x_4$ is not compactified at all. Decompactifying the $x_4$ direction is achieved technically by simply substituting one instead of $f(u)$ in (\ref{S_S_metric}). This model was studied in \cite{Antonyan:2006vw}, and the profile of the probe branes was found to be given by (\ref{u_Lamabda_ne_u}) only with $f(u)=1$ and that the integral could be brought to the closed form
\be
x_4(u) =\frac{1}{8} \bigg(\frac{R^{3}}{u_0}\bigg)^{1/2}[B(\frac{9}{16},\frac12) - B(\frac{u_0^8}{u^8};\frac{9}{16},\frac12)]
\ee
where $B(p,q)$ and $B(x,p,q)$ are the complete and in complete beta functions. The asymptotic separation between the $D8$ and $\bar{D}8$ branes $L$ is given by
\be
L= \frac{1}{4}\bigg(\frac{R^{3}}{u_0}\bigg)^{1/2}B(\frac{9}{16},\frac12)
\ee

 In terms of the dual field theory  the model is  in fact physically  very different from the Sakai Sugimoto model. It is a gravity model dual to a  non-confining gauge theory. Recall that the holographic expression of the string tension is given by (\ref{stringtension}) evaluated at the minimum value of $u$ \cite{Kinar:1998vq} which for the present case is $u=0$ and hence the string tension vanishes.

 The effective five dimensional flavored gauge theory   for the present case is identical to that of the Sakai Sugimoto model,
 namely, the characterization given in (\ref{SS_charactization}) applies also for the uncompactified model with the difference that now the function $\gamma$ is given by
 \be
\gamma= \frac{u^8}{u^8-u_0^8}
 \ee
The action of the gauge fields could be rescaled into the form
\begin{eqnarray}\label{gaug_DBI_D4D8_y}
S_{F^2}=-\frac{T_8(2\pi \alpha')^2 V_{S_4}R^{9/2}_{D4}u_0^{1/2}}{4g_s}
\int d^4xdy\bigg{\{}\bigg{(}\frac{\gamma}{y}\bigg{)}^{1/2}F_{\mu\nu}^2+2\frac{u_0}{R_{D4}^3}\frac{y^{5/2}}
{\gamma^{1/2}}F_{\mu y}^2\bigg{\}}
\end{eqnarray}

Unlike the situation in the compactified case of the previous section, in the AHJK model there is no
compactification scale below which the theory becomes effectively
4d, so the definition of the 't Hooft coupling of the 4d gauge theory is not
so clear. Nevertheless, there are two scales in the problem that could be used to
construct the 4d 't Hooft coupling; $L$ and $\ell_s$.
In the first case we set
\be
\lambda_4=\lambda_{TC}=\frac{\lambda_5}{L}=\frac{g_sN_{TC}\ell_s}{L}
\ee
where $\lambda_5 $
is the 't Hooft coupling of the five dimensional gauge theory.
The asymptotic distance between the $D8$ and $\bar{D}8$ is given numerically by
\be
L=\approx 0.72\frac{R^{3/2}_{D4}}{u_0^{1/2}}
\ee
and we find:
\be
\kappa_8=\frac{T_8(2\pi \alpha')^2 V_{S_4}R^{9/2}_{D4}u_0^{1/2}}{4g_s}=\frac{0.72T_8(2\pi \alpha')^2 V_{S_4}R^6_{D4}}{g_sL}
=\frac{0.48\ell_s g_sN_{TC}^2}{(2\pi)^4L}=\frac{0.48\lambda_{TC}N_c}{(2\pi)^4}
\ee
Solving numerically the e.q.m of the non-normalizabe mode given in (\ref{EOMpsi}), and substituting it
into (\ref{Holographic_S_in_General}) we find that the S-parameter is given by
\be
S_{AJHK}\approx  19.1\kappa_8=.006 \lambda_{TC}N_{TC}
\ee
and we note that the dependence on $u_0/L$ is now hidden inside the
definition of $\lambda_{TC}$ 
\footnote{Note that the gravity description is only valid for $\lambda_5>> L $.}.\\
Using $\ell_s$ instead, the dimensionless ratio is $\lambda_5/\ell_s$ and we find
\be
\tilde{\lambda}_{TC}=\lambda_5/\ell_s=g_sN_{TC}
\ee
so
\be
\kappa_8=
\frac{g_s^{1/2}N_c^{3/2}\sqrt{\pi}}{6(2\pi)^3}\bigg(\frac{u_0}{\ell_s}\bigg)^{1/2}
=\frac{\sqrt{\pi}\tilde{\lambda}_{TC}^{1/2}N_{TC}}{6(2\pi)^3}\bigg(\frac{u_0}{\ell_s}\bigg)^{1/2}
=0.0012\bigg(\frac{u_0}{\ell_s}\bigg)^{1/2}\tilde{\lambda}_{TC}^{1/2}N_{TC}
\ee
And we find that the S-parameter is given by (for $\frac{u_0}{\ell_s}=1$)
\be
S_{AJHK}= .023 \tilde{\lambda}_{TC}^{1/2} N_{TC}
\ee

\section{Non critical $AdS_6$ model}\label{section3}
In the model discussed in the previous section and in  all the models we will encounter
 in the following sections  the flavor branes were wrapping certain  non-trivial cycles
on top of spanning the Minkovski space and the radial direction. In fact the wrapped dimensions have not played any role in the techincolor scenario and in particular in the determination of the S parameter.  This naturally calls for models without the wrapped cycle and in general with as less as possible extra dimensions. Models of this kind are the non-critical gravitational models.
Such a model that may serve as a non-critical dual  of QCD was proposed in
\cite{Casero:2005se}.
This model is based on a non-critical SUGRA  background presented in
\cite{Kuperstein:2004yf},\cite{Kuperstein:2004yk}
which  can be viewed as the backreaction of $N_{TC}$ coincident $D4$ branes
placed in flat 6d space with linear dilaton. In \cite{Casero:2005se}
the model was modified by introducing fundamental quarks via
$N_{TF}$ $D4$ and anti-$D4$  prob branes.

The various fields in the background are:
\begin{eqnarray}\label{Non_crit_background}
ds^2_6=\bigg{(}\frac{u}{R_{AdS}}\bigg{)}^2dx^2_{1,3}+
\bigg{(}\frac{R_{AdS}}{u}\bigg{)}^2\frac{du^2}{f(u)}+
\bigg{(}\frac{u}{R_{AdS}}\bigg{)}^2f(u)dx_4^2
\end{eqnarray}
\begin{eqnarray}
F_{(6)}=Q_c\bigg{(}\frac{u}{R_{AdS}}\bigg{)}^4dx_0\wedge dx_1\wedge dx_2\wedge dx_3\wedge du\wedge dx_4
\end{eqnarray}
\begin{eqnarray}
e^{\phi}=\frac{2}{\sqrt{3}{Q_c}}\ \ \  ; \ \ \
f=1-\bigg{(}\frac{u_{\Lambda}}{u}\bigg{)}^5\ \ \  ; \ \ \
R_{AdS}^2=\frac{15}{2}\ell^2_s
\end{eqnarray}
where $Q_c=\frac{N_{TC}}{2\pi}$ with $N_{TC}$ being the number of $D4$ brans,
$x_4$ is taken to be periodic where to avoid conical singularity
its periodicity is set to
\begin{eqnarray}
x_4\sim x_4+\delta x_4 \ \ \ ;\ \ \ \delta x_4 =\frac{4\pi R_{AdS}^2}{5u_{\Lambda}}
\end{eqnarray}
We also define $M_{\Lambda}$ as a typical scale below which the theory is
effectively four dimensional:
\begin{eqnarray}
M_{\Lambda}=\frac{2\pi}{\delta x_4}=
\frac{5u_{\Lambda}}{2R_{AdS}^2}=
\frac{u_{\Lambda}}{3}
\end{eqnarray}
One should note that unlike in critical SUGRA models, in this model $R_{AdS}$ is a constant independent of $g_s N_{TC}$, hence the curvature is always of order one and there is no way to go to a small curvature regime.

\begin{figure}
\centerline{
\begin{tabular}{cc}
\includegraphics[width=5in]{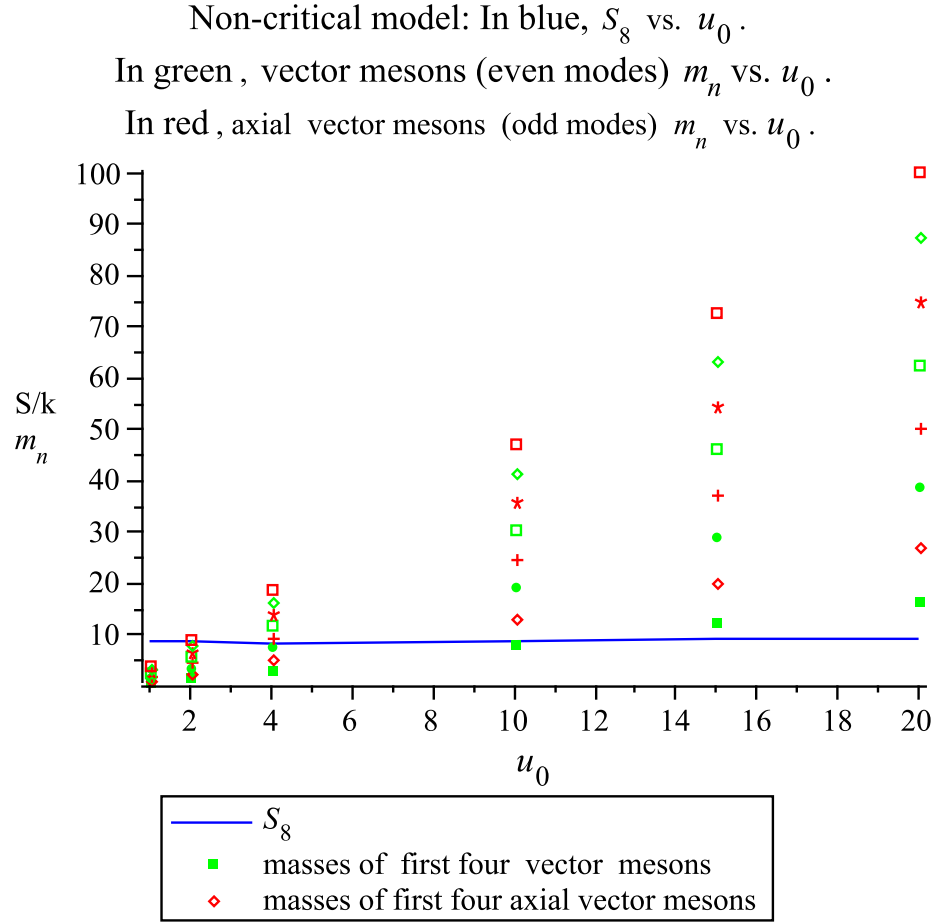}&
\\
\end{tabular}
}
\caption{$S_8$ in the non critical model vs. $u_0$.
The linearity of the vector (axial) mesons masses in $u_0$ could be
seen from their doted green (red) plots for the first 8 modes
 $(u_{\Lambda}=1)$.}
\label{Non_crit_S_8}
\end{figure}

Into this background a set of  $N_{TF}$ pairs of $D4\!-\!\bar{D}4$ prob branes
are placed in a similar manner as in the
Sakai-Sugimoto model, namely they span asymptotically the coordinates $(x_0, ...x_3,x_5)$.
In the corresponding brane configuration, one cannot separate the color and flavor branes, namely the strings that connect the two types of branes are necessarily of zero length, hence it is dual to a field theory system with chiral symmetry.
The profile of the probe branes  is determined by solving the equations of motion that follow from their DBI action. Unlike the critical case, here there is  a priori  an  additional CS term on top of the  DBI action (\ref{DBICS}) of the form  $ S_{CS} = T_4\int {\cal P} ( C_{(5)})$.
However, for reasons given in \cite{Mazu:2007tp}  including the CS term yields
unphysical results,  hence from here on we shall set the CS term to zero.
Thus using only the DBI action the profile of the probe branes is found to be

\begin{eqnarray}
x_{4,cl}(u)=\int_{u_0}^u
\frac{(u_0^5f^{1/2}(u_0))du'}{(\frac{u'}{R_{AdS}})^2f(u')
\sqrt{u'^{10}f(u')-u_0^{10}f(u_0)}}
\end{eqnarray}
Again for convenience we define
\begin{eqnarray}
\gamma=\frac{u^8}{u^{10}f(u)-u_0^{10}f(u_0)}
\end{eqnarray}
 We also rescale $u$ to set $u_{\Lambda}=1$.\\
In terms of the general discussion of section  (\ref{ThSp}), the model is characterized by
\bea
\kappa_{nc}&=& T^{(nc)}_4 ( 2\pi \alpha')^2e^{-\phi}R_{AdS}  = \sqrt{\frac{5}{2}}\frac{3N_{TC}}{16\pi^3}\sim 0.0095N_{TC} \CR
a(u)&=& \gamma^{\frac12} \qquad b(u)=R_{AdS}^{-4}\frac{u^2}{\gamma^{1/2}} \CR
\eea
Once we obtained the solution for the non-normalizable mode by numerical
integration of (\ref{EOMpsi})
we plug it into the holographic definition of S (\ref{Holographic_S_in_General}),
and got an estimation of S,
for the antipodal configuration $u_0=u_{\Lambda}=1$:
\begin{eqnarray}\label{S_nc}
S=10.3\kappa_{nc}=0.095N_{TC}
\end{eqnarray}
The dependence of $S$ on $u_0$ for
$u_0>u_{\Lambda}=1$ is drawn in figure (\ref{Non_crit_S_8}).
It is obvious from the figure that at large $u_0$ $S$ is a constant independent of $u_0$. The asymptotic value it takes is $\frac{S}{\kappa_{nc}}\simeq 6.54$. This behavior will be derived below also qualitatively.

Next we would like to compute the $S$ parameter using the sum rule formula of (\ref{sum_rule}).
To compute $S_8$, the sum over first eight resonance, we need on top of the low lying masses also the corresponding
decay constants. These are determined by solving numerically (\ref{decayconst}).
Substituting the values of the masses and of the decay constants  into (\ref{sum_rule}) and summing  up to $n=8$ we  find
\begin{eqnarray}\label{nc_S_8}
S_8=8.96\kappa_{nc}=0.086N_{TC}
\end{eqnarray}
According to \cite{Hirn:2007bb} it was anticipated that the
higher KK modes will decouple from the spectrum and that $S_n$
for some finite $n$ will produce a good approximation for $S$.\\

For the general case $u_0>u_{\Lambda}=1$, the S-parameter seems
to be almost independent of $u_0$ as could be seen in figure (\ref{Non_crit_S_8}).

In order to see the S-parameter dependence on $u_0$ in a more
qualitative manner we repeat the scaling argument we used
in section (\ref{SS}).
By changing to the dimensionless variable
$y=\frac{u}{u_0}$ we find that after taking the limit $u_0\gg1$
 eq. (\ref{EOMpsi}) and (\ref{decayconst}) becomes
\begin{eqnarray}\label{eom_for_nc_dim_less_psi_r}
\tilde{\gamma}^{-1/2}\partial_y(y^2\tilde{\gamma}^{-1/2}\partial_y\tilde{\psi}_{n}(y))
=-\frac{m_n^2R^4_{AdS}}{u_0^2}\tilde{\psi}_n(y)
\end{eqnarray}
\begin{eqnarray}\label{a_Vn_nc_dim_less}
-\kappa_{nc}(R^4_{AdS})^{-1}y^2
\tilde{\gamma}^{-1/2}\partial_y\tilde{\psi}_{Vn}|_{y=\infty}
=\frac{g_{Vn}}{u_0^2}
\end{eqnarray}
where we noted that in this limit
\begin{eqnarray}
\gamma\to \tilde{\gamma}(y)=\frac{y^8/u_0^2}{y^{10}-1}
\end{eqnarray}

Both in (\ref{eom_for_nc_dim_less_psi_r}) and (\ref{a_Vn_nc_dim_less})
the left hand side is independent of $u_0$,
so the right hand is independent of it as well, and we find
\begin{eqnarray}\label{qualitative behavior of the mesons mass}
m_{Vn/An}^2 \sim \frac{u_0^2}{R^4_{AdS}}
\end{eqnarray}
\begin{eqnarray}\label{qualitative behavior of the mesons mass}
g_{Vn/An} \sim \frac{u_0^2}{R^4_{AdS}}
\end{eqnarray}
Plugging these into (\ref{sum_rule}) we see that indeed the S-parameter
is independent  of $u_0$ in the limit $u_0>>1$.

As was shown in the previous section for the Sakai Sugimoto model, we still need to determine the compactification scale of the system $M_{\Lambda}$. As before taking  the techinipion decay constant to equal the electro-weak scale, we find using the numerical integration

\begin{eqnarray}\label{nc_M_kk}
(246\textrm{GeV})^2= \Pi_A(0)=F_{\pi}^2=
-\kappa_{nc}R_{AdS}^{-4}u^2\gamma^{-1/2}\partial_u\psi_{A}^0(u,0)|_{u=\infty}=0.22\kappa_{nc}M_\Lambda^2
\end{eqnarray}
and the corresponding  mass scale is
\begin{eqnarray}
M_\Lambda^2=\frac{(246\textrm{GeV})^2}{0.0022N_{TC}}
\end{eqnarray}
For $N_{TC}=4$  the scale is found to be $2.4\textrm{TeV}$.
Using this scale combined with the eigenvalues of  (\ref{EOMpsi})
yields the following masses for the first two resonance:
\begin{eqnarray}\label{rho_and_a_mass_nc}
m_{\rho_T}=1.79\textrm{TeV}\ \ \ ;\ \ \ m_{a_{1T}}=2.98\textrm{TeV}
\end{eqnarray}
\\

\section{The KMMW model  with $D6$ and anti-$D6$ flavor branes}\label{section4}
As was emphasized in the last two sections, to have a chiral flavor symmetry of the
form $U_L(N_{TF})\times U_R(N_{TF})$ one has to place a set of $N_{TF}$ probe branes and anti-branes
in such a way that the strings that stretch between them and between the original technicolor branes that constitute the background cannot have a non-trivial length. That required $D8$ and $\bar D8$ branes in the critical model and $D4$ and anti-$D4$ in the non-critical model.
However, to play the role of technicolor one may use a non chiral setup where a symmetry of two sets of Dirac fermions of the  form $U_1(N_{TF})\times U_2(N_{TF})$ is spontaneously broken to a diagonal symmetry $U_D(N_{TF})$.  The dual of such a field theory can be realized  by placing a stack of $D6$ branes and $\bar D6$ branes into Witten's model \cite{Witten:1998zw}.
The construction of the $D4/D6\!-\!\bar{D}6$ system is
identical to that of the Sakai Sugimoto model but with
$D6\!-\!\bar{D}6$ prob branes instead of the
$D8\!-\!\bar{D}8$.
The profile of the probe branes is determined by solving the equations of motion
 for the three coordinates transverse to the branes.  This was done by Kruczenski et al in \cite{Kruczenski:2003uq}. The $D6$ branes span  the ordinary  space-time coordinates,
wrap an $S_2$ inside the $S_4$ and curve along the cigar spanned by $(u,x_4)$ coordinates.
Obviously all the parameters of the background are those of \cite{Witten:1998zw} as was described in section (\ref{SS}).
On the other hand, the induced metric is now
\begin{eqnarray}\label{S_S_metric}
ds^2\!\!=\!\!\bigg( \frac{u}{R_{D4}}\bigg)^{3/2}\!\!\!\!\![\! -dt^2\!\!
+\!\delta_{ij}dx^idx^j]\!+\!R_{D4}^{3/2}u^{1/2}\!d\Omega_2^2
\!+\!\!\bigg[\!\bigg( \frac{u}{R_{D4}}\bigg)^{\!\!3/2}\!\!\!\!\!f(u)(\partial_u x_4)^2\!+\!\!
\bigg{(}\frac{R_{D4}}{u}\bigg)^{\!\!3/2}\!\!\!\!\!\! \frac{1}{f(u)}\bigg{]}du^2
\end{eqnarray}
where we  still sets $u_{\Lambda}=1$.
The curve of the $D6-\!\bar{D}6$ brane on the cigar spanned by $(u,x_4)$
is found via the DBI action to be
\begin{eqnarray}\label{x_u_D4D6}
x_4(u)=u_0^{7/2}f(u_0)^{1/2}\int_{u_0}^udx\frac{1}{x^{3/2}f(x)\sqrt{x^7f(x)-u_0^7f(u_0)}}
\end{eqnarray}

In the terminology of section (\ref{ThSp}) this model is characterized by
\bea
\kappa_{nc}&=& -\frac{{T}_6 ( 2\pi \alpha')^2V_2R^3_{D4} }{g_s}=\frac{N_{TC}}{(2\pi)^2}\approx.025N_{TC}  \CR
a(u)&=& \frac{\gamma^{\frac12} }{u} \qquad b(u)=\frac{u^2}{R_{D4}^3\gamma^{1/2}} \CR
\eea
where
\begin{eqnarray}
\gamma(u)=\frac{u^7}{u^7f(u)-u_0^7f(u_0)}
\end{eqnarray}

Integrating Numerically the equations of motion for the non-normalizable mode we find
that for the antipodal configuration $u_0=u_{\Lambda}=1$
\begin{eqnarray}\label{definition_nc}
S=4\pi\frac{d}{dq^2}(a_V^0(q^2)-a_A^0(q^2))|_{q^2=0}\approx
20.3\kappa_{6}=20.3\times .025N_{TC}=.5N_{TC}
\end{eqnarray}
For the general case $u_0>u_{\Lambda}=1$, we find a slow decrease of S
towards the asymptotical value $S\simeq 18.49$ so S virtually independent of $u_0$.

Now, we want to estimate the S-parameter using the sum over hadronic resonance
given in (\ref{sum_rule}) and see its agreement with (\ref{definition_nc}).
This requires the values of the decay constants of each of the vector and axial-vectoes mesons,
and these are given as in (\ref{decay_const_D4D8}) by:

\begin{eqnarray}\label{decay_const_D4D6}
g_{Vn}=m^2_{Vn}a_{Vn}=-\kappa_6R^{-3}u^2\gamma^{-1/2}\partial_u\psi_{Vn}|_{u=\infty}
\end{eqnarray}
\begin{eqnarray}
g_{An}=m^2_{An}a_{An}=-\kappa_6R^{-3}u^2\gamma^{-1/2}\partial_u\psi_{An}|_{u=\infty}
\end{eqnarray}
We plugged this into (\ref{sum_rule}) and summed up to $n=8$ and found
\begin{eqnarray}
S_8=8.76\kappa_{6}=.194N_{TC}
\end{eqnarray}
We see that as in the Sakai-Sugimoto model, the higher KK modes doesn't
decouple from the spectrum and $S_n$ for some finite $n$ doesn't
produce a good approximation for $S$.\\

We repeat the scaling argument to determine
 qualitatively the dependance of the S-parameter
on $u_0$ we. Changing to the dimensionless variable
$y=\frac{u}{u_0}$ in eq. (\ref{EOMpsi}) and
(\ref{decay_const_D4D6}), then in the limit $u_0>>1$ these become
\begin{eqnarray}\label{eom_for_psi_D4D6_scaled}
y\tilde{\gamma}^{-1/2}\partial_y(y^2\tilde{\gamma}^{-1/2}\partial_y
\tilde{\psi}_{n}(y))=-\frac{m_n^2R^3}{u_0}\tilde{\psi}_{n}(y)
\end{eqnarray}
\begin{eqnarray}\label{decay_const_D4D7_scaled}
-\kappa_6R^{-3}u^2\tilde{\gamma}^{-1/2}\partial_u\psi_{Vn}|_{u=\infty}=\frac{g_{Vn}}{u_0}
\end{eqnarray}
where we denoted
\begin{eqnarray}
\gamma\to \tilde{\gamma}(y)=\frac{y^7}{y^7-1}
\end{eqnarray}

Both in (\ref{eom_for_psi_D4D6_scaled}) and (\ref{decay_const_D4D7_scaled})
the left hand side is independent of $u_0$,
so the right hand is independent of it as well, and we find
\begin{eqnarray}\label{qualitative behavior of the mesons mass}
m_{Vn/An}^2 \sim \frac{u_0}{R^3}
\end{eqnarray}
\begin{eqnarray}\label{qualitative behavior of the mesons mass}
g_{Vn/An} \sim \frac{u_0}{R^3}
\end{eqnarray}
A brief look at (\ref{sum_rule}) tells as that at the limit $u_0\gg1$,
the S-parameter will exhibit independency of  $u_0$.\\
As in the previous cases we determine the compactification scale of the model
by equating the technipion decay constant
to the electro-weak scale. Using numerical integration we find
\begin{eqnarray}
\Pi_A(0)=F_{\pi}^2=
-\kappa_6(R)^{-3}u^2\gamma^{-1/2}\partial_u\psi_{A}^0(u)|_{u=\infty}= 0.47M_{\Lambda}^2\kappa_{6}=(246\textrm{GeV})^2
\end{eqnarray}
For $N_{TC}$ this gives $M_{KK}=1.1\textrm{TeV}$.
Last we add that this mass scale set the masses of the first two
resonance to be: $m_{\rho}=0.7\textrm{TeV}$ , $m_{a_1}=1.16\textrm{TeV}$.

\section{$D5$ branes compactified on two circles  with $D7-\bar D7$ flavor branes}\label{section5}
Another interesting model in the context of HQCD is given by
the near horizon limit of the non-extremal background of $N_c$ $D5$ branes.
Now, adding  $N_{TF}$ $D7\!-\!\bar{D}7$ prob branes into this
background we get open strings between the $D5$ to the $D7$ which are
fundamentals of the $SU(N_c)$ gauge group in doublets of $SU(N_{TF})$.
The fields in this background are given by 
\begin{eqnarray}
ds^2=\frac{u}{R}(\eta_{\mu\nu}dx_{\mu}dx_{\nu}+dx^2_4+f(u,u_{\Lambda})dx_5^2)
+\frac{R}{u}\frac{du^2}{f(u,u_{\Lambda})}+Rud\Omega_3^2
\end{eqnarray}
where
\begin{eqnarray}
f(u,u_{\Lambda})=\bigg{(}1-\frac{u_{\Lambda}^2}{u^2}\bigg{)}\ \ \  ; \ \ \  R^2=g_sN_c\alpha'
\end{eqnarray}
and the dilaton and 3 form field strength are given by
\begin{eqnarray}
\exp(\phi)=g_s\frac{u}{R}\ \ \ ;\ \ \ F_3=\frac{2R^2}{g_s}\Omega_3
\end{eqnarray}
In this model $x_4$ and $x_5$ are compact
\begin{eqnarray}
x_4=x_4+2\pi R_{x_4} \ \ \ ;\ \ \ x_5=x_5+2\pi R_{x_5}
\end{eqnarray}
where in order to avoid conical singularity we must set $R_{x_5}=R$.
We choose the $D7\!-\!\bar{D}7$ prob branes to be space filling and flat
in the $M_4$ and $S_3$ directions and curves on the $(u,x_4,x_5)$ space.
We choose to parameterize the curve by the $u$ coordinate
$(x_4(u),x_5(u),u)$ where the functions $x_4(u)$ and $x_5(u)$
 will be determine by the minimization of the DBI action
\begin{eqnarray}
S_{D7}=\frac{T_7V_3V_4}{g_s}\int duu^3\sqrt{(\partial_ux_4)^2+f(u,u_{\Lambda})
(\partial_ux_5)^2+\frac{R^2}{u^2f(u,u_{\Lambda})}}
\end{eqnarray}
and we find (from here on we will set $u_{\Lambda}=1$)
\begin{eqnarray}
x_4(u)=P_4R\int_{u_0}^u\frac{du'}{\sqrt{u'^2f(u')\bigg{(}u'^6-P^2_4-\frac{P_5^2}{f(u')}\bigg{)}}}
\end{eqnarray}
\begin{eqnarray}
x_5(u)=P_5R\int_{u_0}^u\frac{du'}{\sqrt{u'^2f(u')^3\bigg{(}u'^6-P^2_4-\frac{P_5^2}{f(u')}\bigg{)}}}
\end{eqnarray}
For detailed description of the solutions as a function of the integration constants $P_4$ and $P_5$see \cite{Burrington:2007qd}.
The induced metric is therefore
\begin{eqnarray}\label{S_S_metric}
ds^2\!\!\!\!&=&\!\!\!\bigg( \frac{u}{R}\bigg)\bigg[\eta_{\mu\nu}dx^{\mu}dx^{\nu}+\bigg{(}(\partial_u x_4(u))^2+f(u,u_{\Lambda})(\partial_ux_4(u))^2
+\frac{R^2}{u^2 f(u,u_{\Lambda})}\bigg{)}du^2
+R^2d\Omega_3^2 \bigg]\nonumber  \\
&=&\bigg( \frac{u}{R}\bigg)\bigg[\eta_{\mu\nu}dx^{\mu}dx^{\nu}+
\frac{R^2}{u^2}\gamma(u)du^2
+R^2d\Omega_3^2 \bigg]
\end{eqnarray}
where we defined
\begin{eqnarray}
\gamma(u) \equiv \frac{u^6}{f(u)\bigg{(}u^6-P_4^2-\frac{P_5^2}{f(u)}\bigg{)}}
\end{eqnarray}

The DBI action for the gauge fields on the $D7$ prob branes reads
\begin{eqnarray}\label{D5D7action}
S_{F^2}=-\frac{R^3(2\pi \alpha')^2T_{7}\Omega_3}{4g_s}Tr\int d^4xdu
\bigg{(}\gamma^{1/2}(u)F_{\mu\nu}F^{\mu\nu}+\frac{2u^2}{R^2\gamma^{1/2}(u)}
F_{\mu u}F^{\mu u}\bigg{)}
\end{eqnarray}

Using mode decomposition as in (\ref{decomposition}),
the equation of motion for the $\psi_{n}$ are
\begin{eqnarray}\label{eom_for_psi_r57}
\gamma^{-1/2}\partial_u(u^2\gamma^{-1/2}\partial_u\psi_{n}(u))=-m_n^2R^2\psi_n(u)
\end{eqnarray}
and for the non-normalizable part it is
\begin{eqnarray}\label{eom_for_psi_nr57}
\gamma^{-1/2}\partial_u(u^2\gamma^{-1/2}\partial_u\psi^0(u,q^2))=-q^2R^2\psi^0(u,q^2)
\end{eqnarray}
For the antipodal case $P_4=P_5=0,u_0=1$,   (\ref{eom_for_psi_r57})
and (\ref{eom_for_psi_nr57}) will simplify to
\begin{eqnarray}\label{eom_D5D7}
f^{1/2}\partial_u(u^2f^{1/2}\partial_u\psi_{n}(u))=-m_n^2R^2\psi_n(u)
\end{eqnarray}
and
\begin{eqnarray}\label{eom_D5D7}
f^{1/2}\partial_u(u^2f^{1/2}\partial_u\psi_{n}(u,q^2))=-q^2R^2\psi_n(u,q^2)
\end{eqnarray}
After the change of variable $u^2=1+z^2$, we find
\begin{eqnarray}\label{eqm_z_D5D7_n}
\partial_z\big{(}(1+z^2)\partial_z\psi_{n}(z)\big{)}=-m_n^2R^2\psi_n(z)
\end{eqnarray}
and
\begin{eqnarray}\label{eqm_z_D5D7_nr}
\partial_z\big{(}(1+z^2)\partial_z\psi^0(q^2,z)\big{)}=-q^2R^2\psi^0(q^2,z)
\end{eqnarray}

One can transform (\ref{eqm_z_D5D7_n}) into a standard Schr\"odinger
form via the transformation
\begin{eqnarray}\label{}
z=sinh(x)\ \ \  ;\ \ \ \psi_n(x)=\frac{1}{\sqrt{cosh(x)}}\Psi_n(x)
\end{eqnarray}
and find
\begin{eqnarray}\label{}
\bigg{(}\partial^2_x-\frac{1}{4}\big{(}1+\frac{1}{cosh^2(x)}\big{)}\bigg{)}\psi_{n}=-m_n^2R^2\psi_n
\end{eqnarray}
This eigenvalue problem has only two normalizable solutions and so the sum
in (\ref{sum_rule}) runs only on these two modes.

Substituting (\ref{decomposition}) into (\ref{D5D7action}) we get
$4d$ YM action for the gauge fields with the kinetic terms
canonically normalized provided the $SU(N)$ generator obey
$Tr[T^aT^b]=\frac{1}{2}\delta^{ab}$ and the $\psi_n$ are normalized as
\begin{eqnarray}
\frac{R^3(2\pi \alpha')^2T_{7}\Omega_3}{g_s}\int du \psi_{Vn}\psi_{Vm}
\gamma^{1/2}(u)=\delta_{nm}
\end{eqnarray}
\begin{eqnarray}
\frac{2R^3(\pi \alpha')^2T_{7}\Omega_3}{g_s}\int du\psi_{An}\psi_{Am}
\gamma^{1/2}(u)=\delta_{nm}
\end{eqnarray}
According to our prescription in section (\ref{ThSp}) the boundary terms ,
(\ref{aVn_D4D8}),(\ref{a_V0_D4D8}) will be given by
\begin{eqnarray}
a_{Vn}=-\kappa_7(m_n^2R^2)^{-1}u^2\gamma^{-1/2}
\partial_u\psi_{Vn}|_{u=\infty}
\end{eqnarray}
\begin{eqnarray}
a_{An}=-\kappa_7(m_n^2R^2)^{-1}u^2\gamma^{-1/2}
\partial_u\psi_{An}|_{u=\infty}
\end{eqnarray}
and
\begin{eqnarray}\label{a_0V_D5D7}
a_{V0}=-\kappa_7 R^{-2}u^2\gamma^{-1/2}\partial_u\psi_{V}^0(u,q^2)|_{u=\infty}
\end{eqnarray}
\begin{eqnarray}\label{a_0A_D5D7}
a_{A0}=-\kappa_7 R^{-2}u^2\gamma^{-1/2}\partial_u\psi_{A}^0(u,q^2)|_{u=\infty}
\end{eqnarray}
Where we defined
\begin{eqnarray}
\kappa_7=\frac{u_{\Lambda }R^3(2\pi \alpha')^2\kappa_{7}\Omega_3}{g_s}
\end{eqnarray}
The correlators of the vector and axial-vector currents are given by
the AdS/CFT prescription (\ref{AdS/CFT}), and we find
\begin{eqnarray}\label{AdS/CFT_D5D7}
\Pi_V(q^2)\equiv\langle \mathcal{J}_V^{\mu }(q^2)\mathcal{J}_V^{\nu }(0)\rangle_{F.T}=
-a_{V0}(q^2)
\end{eqnarray}
and
\begin{eqnarray}
\Pi_A(q^2)\equiv\langle \mathcal{J}_A^{\mu }(q^2)\mathcal{J}_A^{\nu }(0)\rangle_{F.T}=
-a_{A0}(q^2)
\end{eqnarray}

\begin{eqnarray}\label{definition_D5D7}
S=-4\pi\frac{d}{dq^2}(\Pi_V-\Pi_A)|_{q^2=0}=-4\pi\frac{d}{dq^2}(a_V^0(q^2)-a_A^0(q^2))|_{q^2=0}
\end{eqnarray}
For the antipodal configuration $(u_0=u_{\Lambda}=1)$
(\ref{a_0V_D5D7}),(\ref{a_0A_D5D7}) becomes

\begin{eqnarray}
a_{0}=-\kappa R^{-2}u^2f^{1/2}\partial_u\psi^0(u,q^2)|_{u=\infty}=
-\kappa(R)^{-2}\frac{u^3}{z}f^{1/2}\partial_z\psi^0(u,q^2)|_{z=\infty}
\end{eqnarray}
using $uf^{1/2}=z$ we find
\begin{eqnarray}\label{asymptotic_a_0}
a_{0}=-\kappa R^{-2}(1+z^2)\partial_z\psi^0(u,q^2)|_{z=\infty}
\end{eqnarray}
The asymptotic behavior of $\psi_{V}^0(u,q^2)$ could be read from
(\ref{eqm_z_D5D7_nr}) by expanding it in powers of $z^{-1}$.
Keeping only the leading order term (\ref{eqm_z_D5D7_nr}) becomes
\begin{eqnarray}\label{}
\partial_z\big{(}z^2\partial_z\psi_{n}\big{)}=-q^2\psi_(q^2,z)
\end{eqnarray}
This has the form of an Euler equation and can be solved using
$\psi(q^2,z)=z^{\alpha_q}$ which leads to an equation for $\alpha_q$
\begin{eqnarray}\label{}
\alpha_q(\alpha_q-1)+2\alpha_q+q^2=\alpha_q^2+\alpha_q+q^2=0
\end{eqnarray}
with the roots
\begin{eqnarray}\label{}
\alpha_q=\frac{-1\pm\sqrt{1-4q^2}}{2}=-\frac{1}{2}\pm\frac{1}{2}\mp q^2
\end{eqnarray}
So we have two asymptotic behavior $\alpha_q^-=-1+q^2$ and $\alpha_q^+=-q^2$.
One correspond to the even mode and one to the odd mode.
Plugging these solution into (\ref{definition_D5D7}) would lead
to a diverging S-parameter!
\newpage
\section{ The conifold model with $D7-\bar D7$ flavor  branes}\label{section6}
So far we have discussed the $S$ parameter of holographic technicolor models that are based on
gravity backgrounds with a cigar like structure of the sub-manifold that includes a coordinate compactified on an $S^1$  circle and the radial direction. This structure of the background ensures the confining nature of
the dual gauge theory and the U shape solutions for the probe brane profile implies the spontaneous breaking of its flavor chiral symmetry.
Recently, another type of a holographic model that admit these two features and which is based on the
conifold geometry was proposed in \cite{Kuperstein:2008cq} and \cite{Dymarsky:2009cm}.
In this section we show that this model also fits the general framework discussed in section (\ref{ThSp})
and we determine the $S$ parameter of the holographic techincolor scenario based on this model.
In fact we can discuss two such models. One based on the conifold geometry  which is a conformal model that does not admit confinement \cite{Kuperstein:2008cq} and a one which relates to the deformed conifold \cite{Dymarsky:2009cm} which is a confining model. To simplify the analysis we discuss here the former but a similar type of calculation can be done also to the latter.
Thus we consider here the conifold background.
The flavor probe brane is taken to be a $D7$ and  $\bar D7$ anti   brane.
It  spans
the space-time coordinates $x_\mu$, the radial direction $u$ and the
three-sphere parameterized by the forms $f_i$ (or alternatively
$w_i$). The transversal space is given by the two-sphere coordinates
$\theta$ and $\phi$. The classical profile depend only on the radial
coordinate.

The $10d$ metric is: \be \label{metric10d:eq} d s_{(10)}^2 =
\frac{u^2}{R_{AdS}^2} d x_\mu d x^\mu + \frac{R_{AdS}^2}{u^2} d s_{(6)}^2
\ee with the $6d$ metric given by
\be\label{metric6:eq}
d s_{(6)}^2=dr^2+\frac{r^2}{3}\bigg(\frac14 (f_1^2+f_2^2)+\frac13f_3^2
+(d\theta-\frac12 f_2)^2+(\sin(\theta)d\phi-\frac12 f_1)^2\bigg)
\ee
and the
$AdS_5$ radius is $R_{AdS}^4=\frac{27}{4} \pi g_s N_{TC}\ell_s^4$. Because the
background has no fluxes except for the $C_4$ form the Chern-Simons
terms do not contribute and the action consists only of the DBI
part \be S_{DBI} \propto\int du
u^3 \left( 1 + \frac{u^2}{6} \left( \theta_u^2 + \sin^2 \theta
\phi_u^2 \right) \right)^{1/2}. \ee Here the subscript $_u$ stands
for the derivatives with respect to $u$.  Setting
$\theta=\pi/2$ we easily find the solution of the equation of
motion \be \label{Solution:eq} \cos \left(
\frac{4}{\sqrt{6}} \phi(u)\right) = \left( \frac{u_0}{u} \right)^4.
\ee There are two branches of solutions for $\phi$ in
(\ref{Solution:eq}) with $\phi \in [-\pi/2,0]$ or $\phi \in
[0,\pi/2]$. For $u_0=0$ we have two fixed ($u$-independent)
solutions at $\phi_-=-\frac{\sqrt{6}}{8} \pi$ and
$\phi_+=\frac{\sqrt{6}}{8} \pi$. The induced $8d$ metric in this
case is that of $AdS_5 \times S^3$ as one can verify by plugging $d
\phi= d \theta=0$ into (\ref{metric6:eq}). For non-zero $u_0$ the
radial coordinate extends from $u=u_0$ (for $\phi=0$) to infinity
(where $\phi(u)$ approaches one of the asymptotic values
$\phi_\pm$). The induced metric has no $AdS_5 \times S^3$ structure
anymore. Notice that unlike the case of the Sakai Sugimoto model, here  the $D7$ probe branes do
not reside at the antipodal points on the $(\theta,\phi)$
two-sphere. This is due to the the  conical
singularity at the tip, so the  $S^2$ does not shrink smoothly.

It is convenient to define a new dimensionless radial
coordinate
\be
z = \frac{u}{R_{AdS}}\sqrt{(1-\frac{u_0^8}{u^8})}
\ee
so that the D7 probe brane stretches along positive $z$ and the anti-brane along negative $z$.

In terms of these the gauge fields action is given by
\be
S_{KW}= \kappa_7 Tr\int d^4 x dz \bigg[\frac{F_{\mu\nu}^2}{\sqrt{z^2+\frac{u_0}{R_{AdS}})^8}} + 16 \bigg(z^2+(\frac{u_0}{R_{AdS}})^8\bigg )^{3/2} F_{\mu z}^2  \bigg ]
\ee
This form of the action translate into the following parameters  in
 the framework for computing the $S$ parameter are
\bea
\kappa_{7}&=& 0.0011N_{TC}  \CR
\hat a(z)&=&\frac{1}{\sqrt{z^2+ u_0^8}} \qquad \hat b(z)=16\bigg(z^2+(\frac{u_0}{R_{AdS}})^8\bigg)^{3/4} \CR
\eea
Repeating the procedure of determining the $S$ parameter using (\ref{Holographic_S_in_General}) we find
\be
S= 0.043 N_{TC}
\ee
and the result using the sum-rule (\ref{sum_rule}) is
\be
S_8=0.036N_{TC}
\ee
For comparison we substitute $N_{TC}=4$ to yield $S=0.17$ and $S_8=0.114$.
Equating as before the technipion decay constant to the electroweak scale
we find that $M_\Lambda$ is given by
\be
M^2_\Lambda = \frac{(246\textrm{GeV})^2}{0.202N_{TC}}
\ee
so that for $N_{TC}=4$ we get $M_\Lambda=1.5\textrm{TeV}$
which gives $m_{\rho}=3.\textrm{TeV}$ , $m_{a1}=5.\textrm{TeV}$.

\section{Summary}\label{section7}
In this paper we have  examined a variety of technicolor
models through their holographic duals.
We have focused mainly on the S-parameter of these models. For that purpose we presented
the method used in \cite{Carone:2007md} to deduce the holographic S-parameter and
showed how to apply the technique to general (suitable) background and then applied it on several models.
Generically Technicolor models admit a confinement behavior and spontaneous flavor chiral symmetry breaking. Indeed some of the  models we have chosen, the Sakai Sugimoto model, the non-critical model
and the model based on $D5$ branes admit both these properties in their low energy regime.
However, we have chosen also other type of models. The uncompactified Sakai Sugimoto model is dual to
a NJL like model. It does not admit confinement but does undergo a spontaneous flavor chiral symmetry breaking.
The conifold model is also non-confining. In fact prior to adding the flavor branes it is invariant under  conformal symmetry  which is spontaneously broken due to the addition of the flavor branes.
The KMMW model with $D6$ branes is confining and it has a symmetry breaking of the form $U(N_{TF})\times U(N_{TF})\rightarrow U_V(N_{TF})$. However, it is not a symmetry of chiral fermions but rather a symmetry of Dirac fermions. From the point of view of the S-parameters there is not much difference between the models that admits both confinement and spontaneous flavor chiral symmetry breaking to the other models.

The direct estimation of the  Peskin-Takeuchi S-parameter  for a strongly
interacting sector is still a grave problem in technicolor model-building. But as was
shown in \cite{Carone:2007md} for the Sakai-Sugimoto model, and also in the present paper
a reliable estimate
for the S-parameter  is with in reach if the field theory has a gravity dual. Strictly speaking the latter applies only for large $N_{TC}$ and large $\lambda_{TC}$.

The results of the S-parameter and the low lying technivector mesons is summarized in table \ref{table_Of_S}.

\begin{table}
\begin{tabular}{|c|c|c|c|c|}
\hline  & & & & \\  & $S(u_0/u_{\Lambda}=1)$ &   $S_8(u_0/u_{\Lambda}=1)$ & $m_{\rho}$ &  $m_{a_1}$
\\
 & & & & \\
\hline\hline
$D4\!-\!D8$ (SS model) & $.017\lambda_{TC}N_{TC}$ &$-.001\lambda_{TC}N_{TC}$ & $\frac{10.6}{\sqrt{\lambda_{TC}N_{TC}}}\textrm{TeV} $ & $\frac{15.6}{\sqrt{\lambda_{TC}N_{TC}}}\textrm{TeV} $ \\ & & & & \\
\hline
$D4-D8$ (AHJK model) & $.006\lambda_{TC}N_{TC}$ & - & -   &- \\ & & & & \\
\hline
$D4-D6$ (KMMW model)& $0.5N_{TC}$ &  $.194N_{TC}$ & $1.4\textrm{TeV}/\textrm{\small{N}}_{TC}^{1/2}  $ &$2.32\textrm{TeV}/\textrm{\small{N}}_{TC}^{1/2}  $\\ & & & & \\
\hline
$AdS_6+D4$  & $.095N_{TC}$ &  $0.086N_{TC}$  & $1.79\textrm{TeV}/\textrm{\small{N}}_{TC}^{1/2} $   &2.98$\textrm{TeV}/\textrm{\small{N}}_{TC}^{1/2}  $\\ & & & & \\
\hline
$D3\!-\!D7$ (KW model+$D7$)&$0.043N_{TC}$ &$0.036N_{TC}$ &$6.\textrm{TeV}/\textrm{\small{N}}_{TC}^{1/2}  $ & $10.\textrm{TeV}/\textrm{\small{N}}_{TC}^{1/2}  $\\ & & & & \\
\hline
$D5-D7$ & $\infty $&  -  &$- $ &$- $\\ & & & & \\
\hline
\end{tabular}\caption{The S-parameter of the six models,
$ N_{TF}=2$, S is given by the AdS/CFT dictionary
(\ref{Holographic_S_in_General}), $S_8$ is the sum over
the first eight modes in (\ref{sum_rule}).}
\label{table_Of_S}
\end{table}

\begin{figure}
\centerline{
\begin{tabular}{cc}
\includegraphics[width=5in]{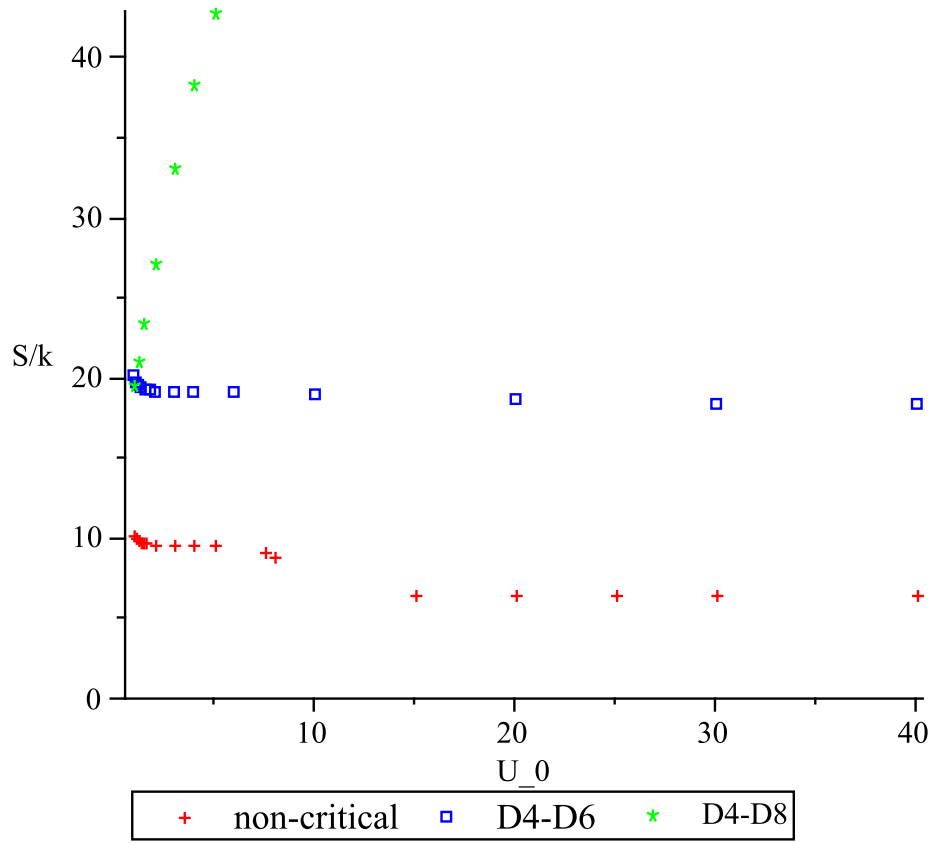}&
\\
\end{tabular}
}
\caption{S-parameter
of the $AdS_6+D4$ non-critical model, $D4-\!D8$, and $D4-\!D6$ models
vs. $u_0$ $(u_{\Lambda}=1)$.}
\label{plot_S_of_u_0}
\end{figure}

In general the S-parameter is a function of all the free parameters of the theory $N_{TC}, \lambda_{TC}, N_{TF}$ and $u_0$ or instead the ``string endpoint" masses defined in (\ref{sepm}).
As for the dependence on $N_{TC}$ and $ \lambda_{TC}$ there is a striking difference between the Sakai-Sugimoto model both the compactified and the uncompactified and the rest of the models.
Whereas in the former models $S$ depends linearly on the product of $N_{TC}\lambda_{TC}$, in the latter models it does not depend on $ \lambda_{TC}$ but rather it is linear only in $N_{TC}$.
The dependence of the S-parameter on $u_0$ in some of the models is drawn in figure \ref{plot_S_of_u_0}.

We can see  from this figure  that while the S-parameter in
Sakai-Sugimoto model (and also its uncompactified cousin)  grows linearly with $u_0$,
 the $D4\!-\!D6$ and $AdS_6+D4$ models exhibit minor dependence on this parameter.
 As explained around (\ref{sepm}) the $u_0$ parameter is related to the string endpoint mass which is given roughly by $\frac{M_m-T_{st}L}{2}$ where $M_m$ is the mass of the corresponding meson, $T_{st}$ is the string tension and $L$ is the length of the stringy meson. This mass parameter has nothing to do with the current algebra or ``QCD" mass and hence one cannot compare it to the dependence on the mass found in \cite{Peskin} at the weak coupling regime.

 The dependence on $N_{TF}$ is more tricky. If one naively embed the $U(2)\in U(N_{TF}$ in such a way that the generator of $SU(2)$ for instance  $T_3$ is just one and -minus one in the upper terms along the diagonal, then there is no dependence of the $S$ parameter on $N_{TF}$ since it relates to the electroweak currents that are affected only by the upper $2\times 2$ block of the $N_{TF}\times N_{TF}$ matrices. However, if we generalize the models discussed in the paper with only a single factor of $SU(N_{TF}=2)$, to a set of $\frac{N_{TF}}{2}$ of such group factors, this should yield an $S$ parameter which is $\frac{N_{TF}}{2}$ times bigger than the one of a single group factor.
 The holographic realization of such a scenario is by taking $\frac{N_{TF}}{2}$ pairs of U shape flavor probe brane and distribute them along the radial direction, namely assign to each of them a different $u_0$. In the non-critical, KMMW and KW$+D7$ models the S- parameter barely depends on $u_0$ and hence a summation over all the pairs of U-shape flavor branes is definitely justified. However for the Skai Sugimoto model and his uncompactified cousin, the S-parameter depends linearly on $u_0$ and thus a naive summation is incorrect. One can of course introduce  very small differences in the values of the $u_0$ associated with each pair and in this way the summation result will be a reasonable approximation.\\
We demonstrate these results for the $AdS_6+D4$ non-critical model and the KMMW model.
For the anti-podal configuration S is given by
\begin{eqnarray}\label{}
S=10.3\kappa_{nc}=10.3\sqrt{\frac{5}{2}}\frac{3N_{TF}N_{TC}}{32\pi^3}
=0.048N_{TF}N_{TC}
\end{eqnarray}
and in the $D4-D6$ system it is
\begin{eqnarray}\label{}
S=20.3\kappa_{6}=20.3\frac{N_{TF}N_{TC}}{2(2\pi)^2}
=.25N_{TF}N_{TC}
\end{eqnarray}
Of course holography is not the only way to estimate the S-parameter,
in \cite{Peskin} a few phenomenological formula where suggested in order to
estimate the S-parameter of technicolor models with QCD like dynamics.
The starting point of their formulas is to use eq. (\ref{sum_rule})
with the masses and decay constants of the techni-hadrons given
by assuming the large-$N$ rescaling relations between
these to their QCD counterparts. 
They found by summing over the first two hadronic resonance $\rho_{TC}$ and $a_{1Tc}$,
that for a model with $SU(N_{TC})$ technicolor gauge group and $N_{TF}$ $SU(2)$
doublets S is given by
\begin{eqnarray}\label{}
S_2\approx 0.247\frac{N_{TF}}{2}\frac{N_{TC}}{3}
\end{eqnarray}
While our holographic summation over the first two resonance gave
\begin{eqnarray}\label{}
S_{2,AdS_6}\approx 0.213\frac{N_{TF}}{2}\frac{N_{TC}}{3}
\end{eqnarray}
Comparing these two estimates reveals a remarkable agreement between
these two very different machineries!

We can continue and compare the estimated mesons masses according to
the large-$N$ scaling relations
\begin{eqnarray}\label{}
m^2_{\rho_{T}}\approx\frac{6}{N_{TC}N_{TF}}\frac{F_{\pi}^2 m^2_{\rho}}{f_{\pi}^2}
\ \ ;\ \
m^2_{a_{1T}}\approx\frac{6}{N_{TC}N_{TF}}\frac{F_{\pi}^2 m^2_{a_1}}{f_{\pi}^2}
\end{eqnarray}
using the data
\begin{eqnarray}\label{}
m_{\rho}\approx 775\textrm{Mev}\ \ ;\ \ m_{a_{1}}\approx 1230\textrm{Mev}
\ \ ;\ \
F_{\pi}^2\approx (246\textrm{GeV})^2\ \ ;\ \ f_{\pi}^2\approx (92\textrm{GeV})^2
\end{eqnarray}
we find for $N_{TC}=4$ and $N_{TF}=2$
\begin{eqnarray}\label{}
m_{\rho_{T}}\approx 1.79\textrm{TeV}
\ \ ;\ \
m_{a_{1T}}\approx 2.8\textrm{TeV}
\end{eqnarray}
By holography we found in (\ref{rho_and_a_mass_nc})
(using (\ref{nc_M_kk}) to set
the Kaluza-Klein scale to $M_\Lambda=2.4\textrm{TeV}$
by which we measure all quantities in the theory)
\begin{eqnarray}\label{}
m_{\rho_{T}}\approx 0.74M_\Lambda=1.78\textrm{TeV}
\ \ ;\ \
m_{a_{1T}}\approx  1.23M_\Lambda=2.98\textrm{TeV}
\end{eqnarray}
Again we find an agreement within a few percent between the two.

\section*{Acknowledgements}

It is a pleasure to thank Ofer  Aharony for very useful
conversations and for his comments on the manuscript.
We are also grateful to  Stanislav Kuperstein
for fruitful discussions.
This work  was supported in part by a centre of excellence supported by the
Israel Science Foundation (grant number 1468/06), by a grant (DIP
H52) of the German Israel Project Cooperation, by a BSF grant,
by the European Network MRTN-CT-2004-512194 and
by European Union Excellence Grant MEXT-CT-2003-509661.

\end{document}